\documentstyle[12pt,epsfig]{article}
\begin{document}

\begin{center}
{\large \bf
A unique parametrization of the shapes of secondary dilepton spectra
observed in central heavy-ion collisions at CERN-SPS energies}\\[6mm]
{\sc
K. Gallmeister$^a$, 
B. K\"ampfer$^a$, 
O.P. Pavlenko$^{a,b}$,
C. Gale$^c$} \\[6mm]
$^a$Forschungszentrum Rossendorf, PF 510119, 01314 Dresden, Germany \\[1mm]
$^b$Institute for Theoretical Physics, 252143 Kiev - 143, Ukraine\\[1mm]
$^c$Physics Department, McGill University, Montreal, QC, H3A 2T8, Canada
\end{center}

\vspace*{9mm}

\centerline{Abstract} 
A unique parametrization of secondary (thermal) dilepton yields in
heavy-ion experiments at CERN-SPS is proposed.
This parametrization resembles a thermal $q \bar q$ annihilation rate.
This is inspired by the observation that lepton pair production rates
are quantitatively similar, whether expressed in a hadronic or partonic
basis. Adding the thermal yield and
the background contributions (hadronic cocktail,
Drell-Yan, correlated semileptonic decays of open charm) 
the spectral shapes of the
CERES/NA45, NA38, NA50 and HELIOS/3 data 
from experiments with lead and sulfur beams can be well
described.\\[3mm]

\section{Introduction} 

The typical temperature scales in heavy-ion collisions
at CERN-SPS energies, extracted from hadron abundances 
\cite{abundances}
and hadron transverse momentum spectra \cite{J.Phys.,Heinz,Cassing}, 
are in the order of 
$T_c$ (or somewhat less), where
$T_c$ stands for the expected deconfinement and/or chiral symmetry
restoration temperature in strongly interacting matter.
For temperatures around that value, 
a critical comparison of dilepton-producing channels suggests that, 
in the intermediate invariant mass region, the sum of various hadronic
processes essentially equals the quark-antiquark annihilation rate into lepton
pairs \cite{LiGale}. 
Similarly, at lower invariant masses the various contributions to the
lepton pair spectrum add up to produce a structure where the $\rho$
meson has acquired a considerable width \cite{RappWambach}. Remarkably,
there also the net dilepton spectrum closely mimics that from a free
gas of annihilating quarks and antiquarks \cite{Klingl_Weise}. 
It is tempting to ascribe this duality of representations to 
a full or partial
restoration of chiral symmetry. This interesting conjecture however 
needs yet to be put on a firm theoretical basis. We simply intent here
to exploit its empirical power. The interested reader
can consult references \cite{Huang,Leonidov_Ruuskanen} for early
analyzes, \cite{RappWambach} for a recent review, and 
\cite{RappShuryak,Phys.Lett.,Schneider} for a recent application. 

We employ here the $q \bar q$ rate 
as a convenient parametrization of the dilepton emissivity
and analyze uniquely 
both the $e^- e^+$ and $\mu^- \mu^+$ channels of the dilepton spectra,
obtained in sulfur and lead beam reactions at the CERN-SPS.
We shall make use of the analytical simplification brought about by
only considering this simple process, to attempt a global fit to the
world data of lepton pair production in central
heavy-ion collisions at SPS 
energies. Since we use the $q \bar q$ rate 
in the full invariant mass range considered and
for the full time evolution of matter as well, 
one could phrase our approach as
a test of an  ''extended duality'' hypothesis. Again, for the moment we
refrain from attempting to formulate a theoretical foundation for this
fact and we first seek to check its validity in a study of heavy-ion
phenomenology.

The dilepton spectra are obtained by a convolution of a rate 
with the space-time evolution of matter. The time evolution is split up
in several stages which can be dealt with separately: 
(i) First, on very short time scales there are hard initial processes
among the partons, being distributed according to primary nuclear
parton distributions, such as the Drell-Yan process
(to leading order $ q \bar q \to l \bar l$)
and charm production
(to leading order mainly $gg \to c \bar c$).
(ii) On intermediate times scales there are the so-called secondary
interactions among the constituents of the hot and dense,
strongly interacting matter. This stage is often denoted as thermal era and
the emitted dileptons as thermal dileptons. 
(iii) If the interactions among the hadrons in a late stage cease, there
are hadronic decays into dileptons and other decay products.

In the present work we focus on a parametrization of the dilepton
yield from stage (ii) with a minimal parameter set 
and assume that the background
contributions from stages (i) and (iii) are known, either by up-scaling
the Drell-Yan and open charm yields from $pp$ to $AB$ collisions or by
using directly the experimentally determined hadronic cocktail.
One should be aware that such a schematic description may suffer from several
deficits, such as missing secondary Drell-Yan processes \cite{Stoecker},
or hadronic final state interaction of open charm \cite{Lin_Wang}
or non-equilibrium effects \cite{PLB_91}.
Also, if the temperature is initially large enough, i.e.\ $T \gg T_c$,
and the matter resides in a deconfined state, one expects a dilepton
spectrum being different from the simple $q \bar q$ rate both in the
low-mass region \cite{Peshier_Thoma} and in the intermediate-mass region
\cite{PRC_95} (cf.\ \cite{Aurenche}).

The aim of the present work is to look for a minimal 
description
of data in a similar spirit as one comfortably parametrizes the transverse
momentum spectra of hadrons by effective slope parameters and afterwards
maps on flow and freeze-out temperature and non-equilibrium components.
One can also draw from experiences with direct photons, where a simple
parametrization of hadronic rates has simplified calculations
considerably \cite{photons}. 

Our paper is organized as follows. In section 2 the model for the dilepton
emission rate is presented. The analyses of the experimental spectra
are performed in section 3. 
Flow effects and time evolution effects are considered in section 4.
The results are summarized in 
section 5. The calculation of the background contributions is
explained in the appendix A.

\section{Model parametrization}

Let $dN/( d^4 x \, d^4 Q)$ be the Lorentz invariant dilepton production rate
from matter in local thermal equilibrium being characterized by a 
temperature $T$,
chemical potentials $\mu_i$, and
four-velocity $u^\nu$ of the flow.
We base the production
rate \cite{LiGale,RappWambach} on the lowest-order quark-antiquark 
annihilation rate
\begin{equation}
\frac{dN}{d^4 Q \, d^4 x} = \frac{5 \alpha^2}{36 \pi^4}
\exp \left\{ - \frac{u \cdot Q}{T} \right\}, 
\end{equation}
where $Q^\mu = (M_\perp \cosh Y, M_\perp \sinh Y, \vec Q_\perp)$
is the lepton pair four-momentum with  transverse mass $M_\perp$, transverse
momentum $\vec Q_\perp$, related to the invariant mass via
$M =\sqrt{M_\perp^2 - Q_\perp^2}$, and rapidity $Y$.
The above rate is in Boltzmann approximation, and a term related to
the chemical potential(s) is suppressed. 
The space-time integration can be performed if a model 
for the set of parameters
$T$, $\mu_i$, $u^\mu$ and their space-time dependence is at disposal.
An obviously strong approximation is to replace all state variables
by averages
\begin{equation} 
\frac{dN}{d^4 Q} = 
\frac{5 \alpha^2}{36 \pi^4}
\exp \left\{ - 
\frac{\langle u \rangle \cdot Q}{\langle T \rangle } \right\}
\,
\int d^4x.
\end{equation}
It is one of the purposes of this paper to investigate the validity or 
the consequences of this potentially extreme, but remarkably simple
assumption.
In section 4 we discuss that the flow effects can be neglected in some
region of the phase space. Then, by assuming a thermal source at
midrapidity $Y_{\rm cms}$, one can make the replacement
$\langle u^\mu \rangle \to (\cosh Y_{\rm cms}, \sinh Y_{\rm cms}, \vec 0)$. 
Taking the space-time volume $\int d^4 x = N_{\rm eff}$ 
as normalization factor 
(which now can also include effects of finite chemical potetials)
one gets the approximation
\begin{equation}
\frac{dN}{d^4 Q} = 
\frac{5 \alpha^2}{36 \pi^4} N_{\rm eff}
\exp \left\{ - 
\frac{M_\perp \cosh (Y - Y_{\rm cms} )}{T_{\rm eff}} \right\},
\end{equation}
where $T_{\rm eff} \equiv \langle T \rangle$. The two parameters
$T_{\rm eff}$ and $N_{\rm eff}$ are to be adjusted to the experimental data.

In a more detailed description 
these two parameters are mapped on a much larger parameter space,
which however is constrained by hadronic observables and allows
a detailed microscopic justification of $T_{\rm eff}$ and $N_{\rm eff}$. 

In what follows we use Eq.~(3) and 
confront it with the data. In doing so we implement the corresponding
detector acceptance, which is most conveniently done by generating the  
six-fold differential rate   
\begin{equation}
\frac{dN}{ p_{\perp 1} \, d p_{\perp 1} \, 
p_{\perp 2} \, d p_{\perp 2} \,
dy_1 \, dy_2 \,
d \phi_1 \, d \phi_2 } 
=
\frac{1}{2 \pi} \frac{dN}{d^4 Q},
\end{equation}
where $p_{\perp 1,2}$, $y_{1,2}$ and $\phi_{1,2}$ denote the transverse
momenta, rapidities and azimuthal angles of the individual leptons 1 and 2,
which must be appropriately combined to construct the pair mass $M$,
the pair transverse momentum $Q_\perp$ and transverse mass $M_\perp$.

\section{Analysis of dilepton spectra} 

\subsection{Lead beam data}

The dilepton experiments in the $e^- e^+$ and $\mu^- \mu^+$ channels 
with the lead beam at CERN-SPS use as targets
either Au or Pb. We neglect the small differences of the target nuclei
and attempt a unique parametrization. The rapidity coverage of the two
experiments is also fairly symmetric around midrapidity
for these symmetric collisions.

\subsubsection{CERES experiment \label{CERES_lead}}

\begin{figure}[t] 
\centering
~\\[-.1cm]
\psfig{file=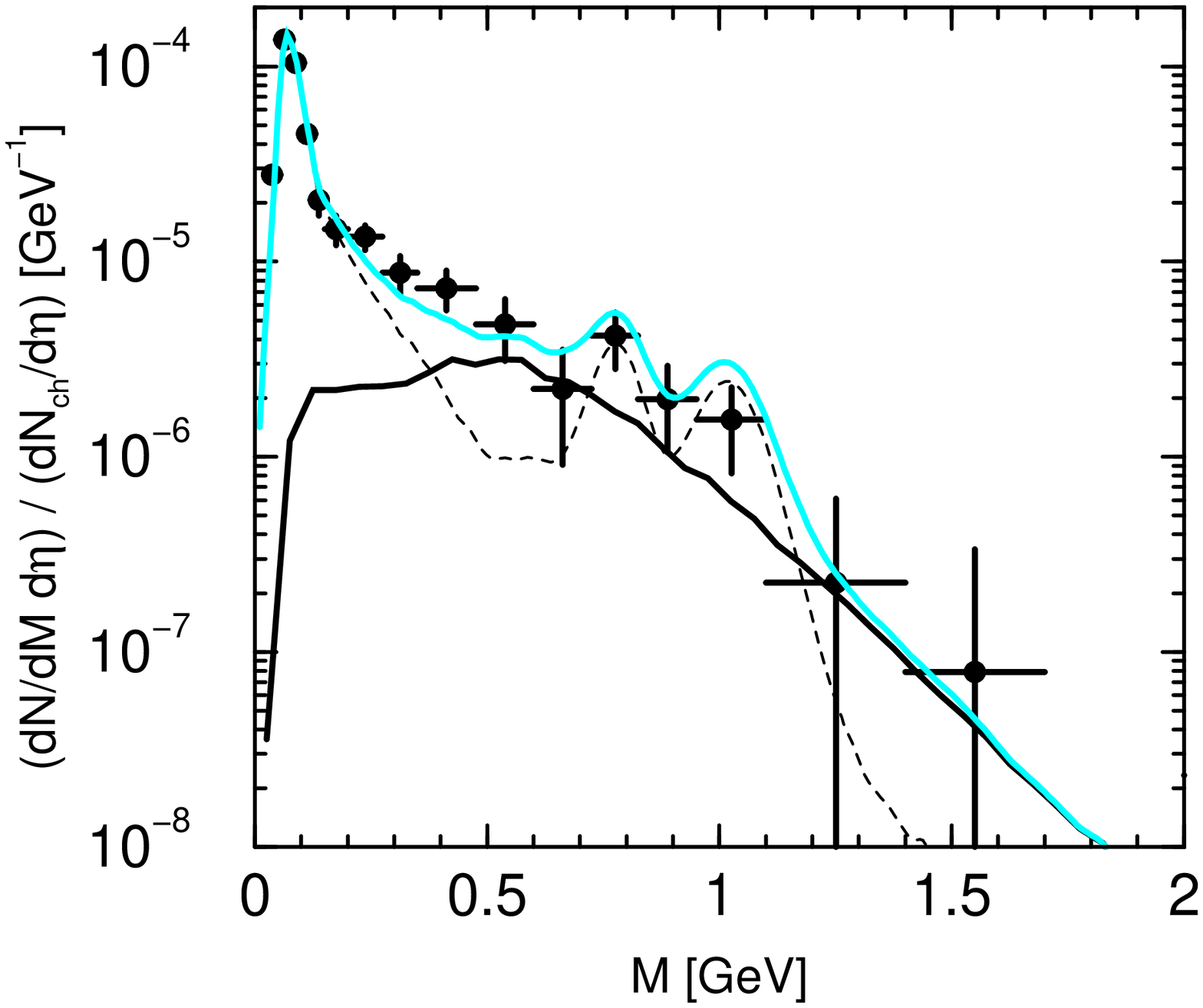,width=6cm,angle=-90}

\psfig{file=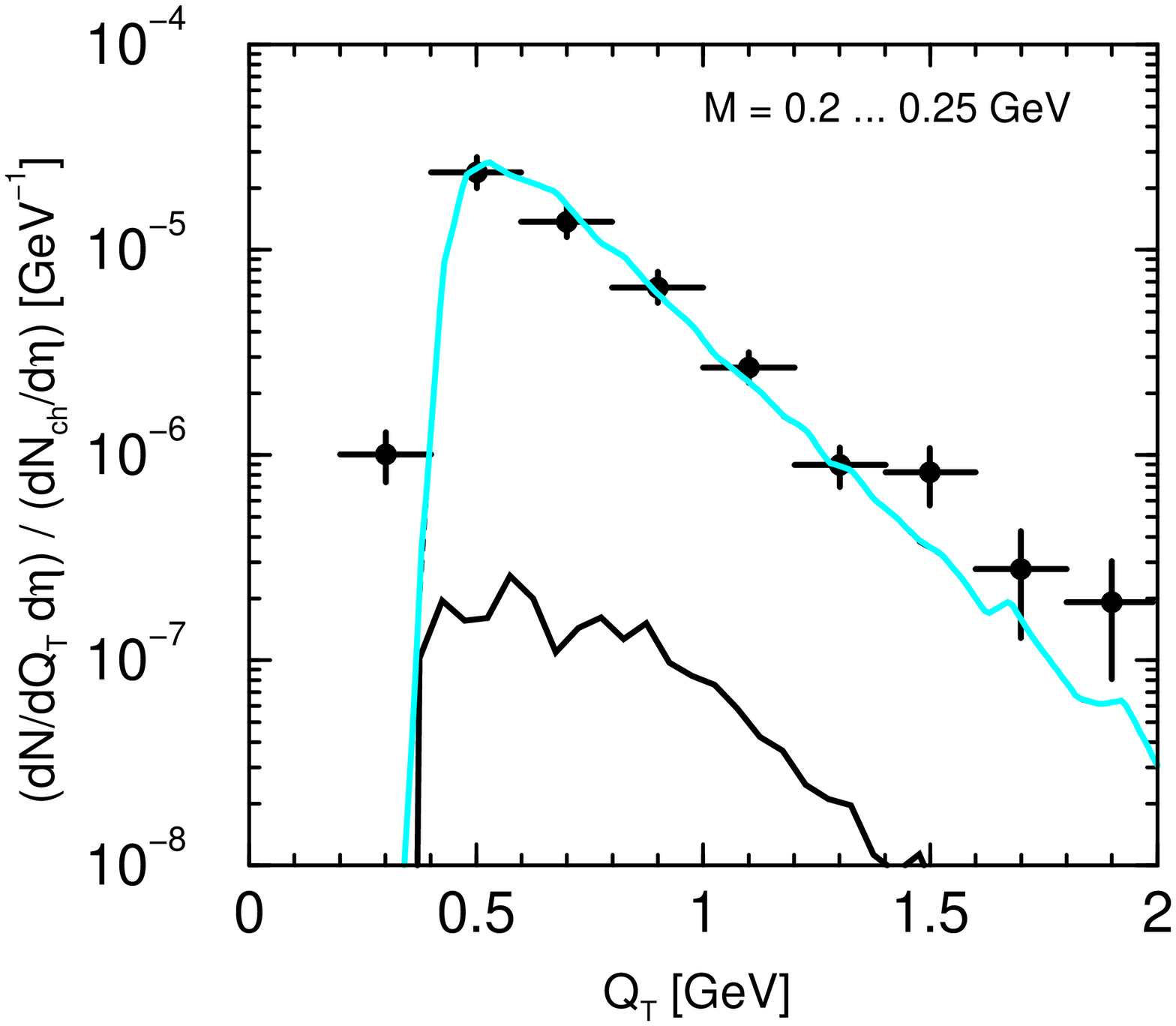,width=6cm,angle=-90}
\hfill
\psfig{file=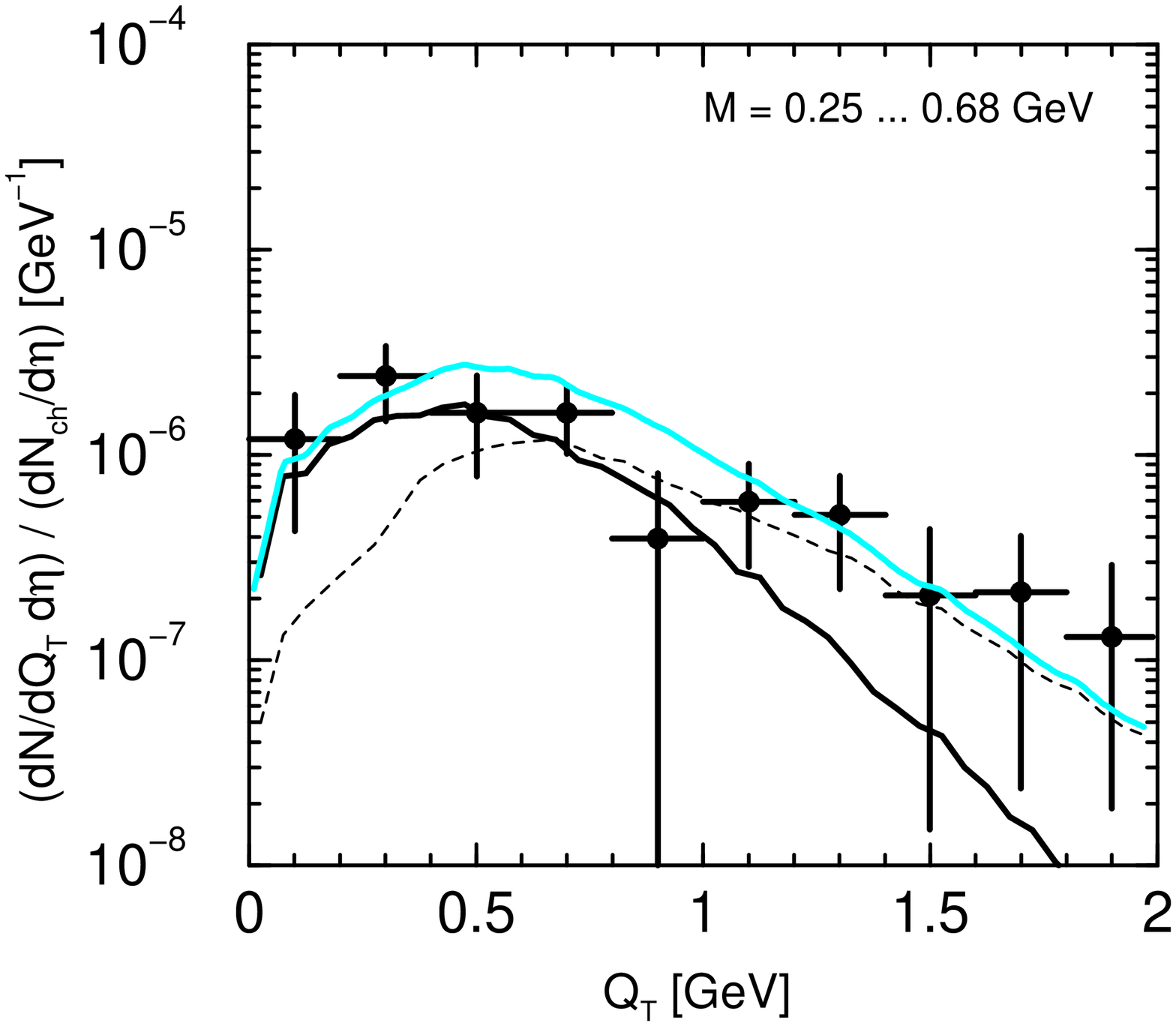,width=6cm,angle=-90}
\hfill
\psfig{file=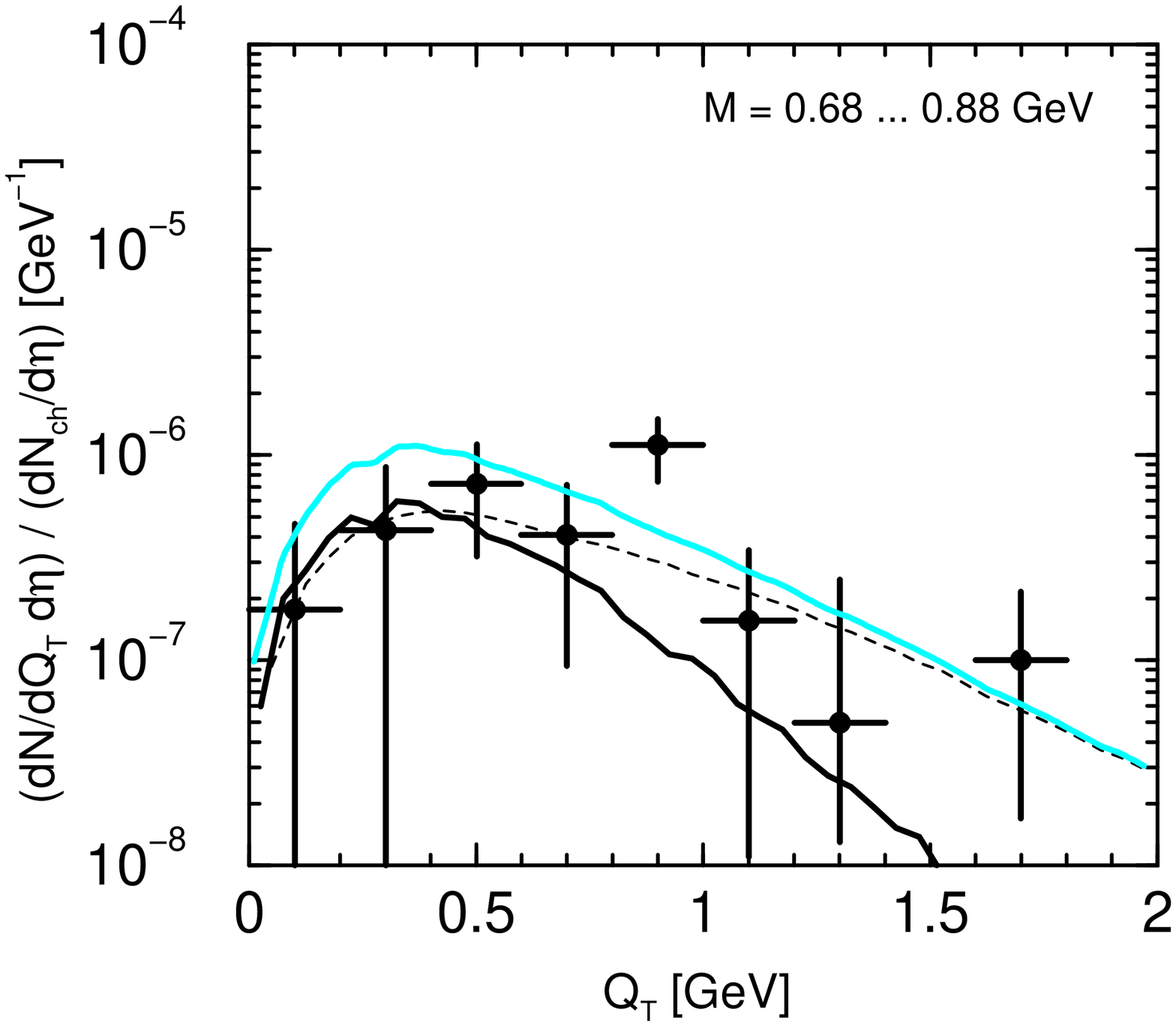,width=6cm,angle=-90}
\hfill
\psfig{file=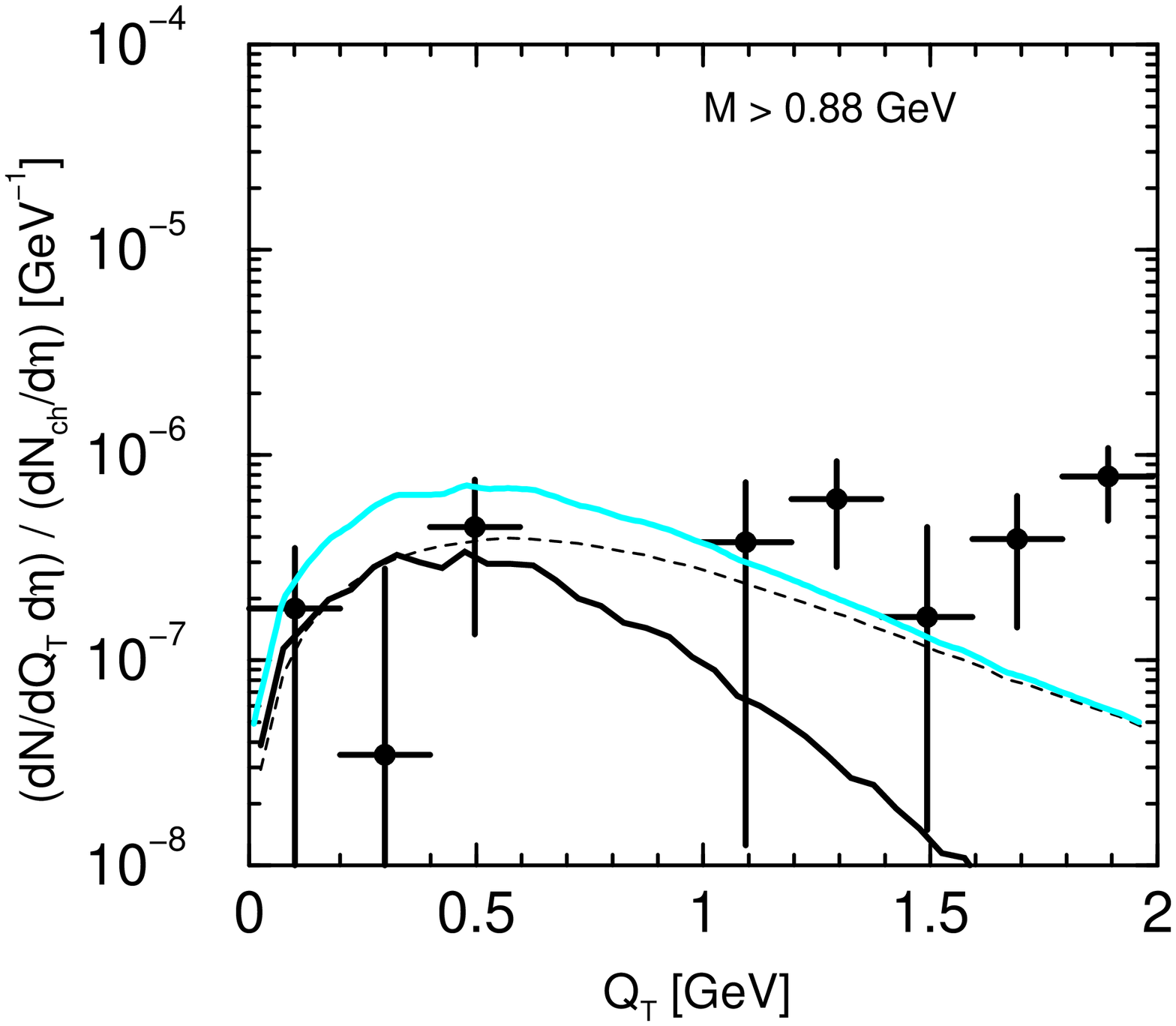,width=6cm,angle=-90}
~\\[.5cm]
\caption{
Comparison of our model with the preliminary 
CERES data \cite{CERES_Pb}. 
Upper middle panel: invariant mass distribution,
lower four panels: $Q_\perp$ spectra for various
invariant mass bins
(dashed lines: hadronic cocktail,
solid curves: thermal yield,
gray curves: sum).
}
\label{f_1}
\end{figure}

A comparison of the model defined by Eqs.~(3, 4) 
with the lead beam data \cite{CERES_Pb}
is displayed in Fig.~\ref{f_1}. With the parameter set of 
$Y_{\rm cms} = 2.9$,
$T_{\rm eff} =$ 170 MeV and 
$N_{\rm eff} = 3.3 \times 10^4$ fm${}^4$ 
a fairly good description of the data
is accomplished. We mention that the use of 
external information, namely
the hadronic decay cocktail and the normalization of
$\langle N_{ch} \rangle = 250$, is essential for describing the CERES
$e^+ e^-$ data \cite{CERES_Pb} in central reactions Pb(158 AGeV) + Au.
The normalization to $\langle N_{ch} \rangle = 250$ corresponds to
a centrality criterion of 
$\sigma_{\rm trigger} / \sigma_{\rm tot} = 0.3$ \cite{CERES_Pb}.

We model the CERES acceptance by $p_{\perp 1,2} >$ 0.2 GeV,
$\eta_{1,2}^{\rm lab} = 2.1 \cdots 2.65$ (with $\eta$ as
pseudo-rapidity) and a relative angle
$\Theta^{\rm lab}_{12} > 35$ mrad between the electron and positron.

In agreement with other models \cite{RappWambach_new,V.Koch}
we also describe the $Q_\perp$ dependence in various $M$ bins
(see Fig.~\ref{f_1}). 

It should be emphasized that
the above parameters $T_{\rm eff}$ and $N_{\rm eff}$ deliver also 
an optimum description of 
the direct photon data \cite{WA98}, as shown in \cite{our_photons}.

\subsubsection{NA50 experiment}

For a description of the NA50 $\mu^+ \mu^-$ data \cite{NA50} 
in central reactions Pb(158 AGeV) + Pb, the Drell-Yan contribution 
and the correlated semileptonic decays of open charm mesons 
are needed.
The latter ones are generated with PYTHIA \cite{PYTHIA}
where the Drell-Yan
K factor of 1.5 is adjusted to the data \cite{DY_K_factor,DY_K_factor_2} 
(cf.\ Appendix \ref{Appendix A1}) and
the open charm K factor of 5.7 to the compilation cross sections
of identified hadronic charm channels \cite{PBM}
and $\mu^+ \mu^-$ data in the reaction
p(450 GeV) + W \cite{pW_Capelli} (cf.\ Appendix \ref{Appendix A2}).

To translate the cross sections delivered by PYTHIA 
into rates we use a thickness function of 31 mb${}^{-1}$
for central collisions Pb + Pb. Actually, however, the centrality criterion
is a selection of data from $E_T$ bin 9 with impact parameter
average $\langle b \rangle < 3.3$ fm and a participant number
$\langle N_{\rm part} \rangle = 381 \pm 7$ according to \cite{NA50}.

The NA50 acceptance is modeled by
$Y^{\rm lab} = 2.9 \cdots 3.9$, a Collins-Soper angle
$\vert \cos \Theta_{\rm CS} \vert < 0.5$ and
the condition of $E > E_{\rm min}$ for the minimum energy of a muon in the
laboratory, where
$E_{\rm min} = E_0 + \Delta E$ with
$E_0 = 11.5$ GeV and
$\Delta E = 16000 (\Theta - 0.065)^2$ GeV,
$0$,
$13000 (\Theta - 0.090)^2$ GeV for
0.037 $< \Theta <$ 0.065,
0.065 $< \Theta <$ 0.090,
0.090 $< \Theta <$ 0.108, respectively. Note that these cuts describe
the NA50 acceptance only approximately.

\begin{figure}[t] 
\centering
~\\[-.1cm]
\psfig{file=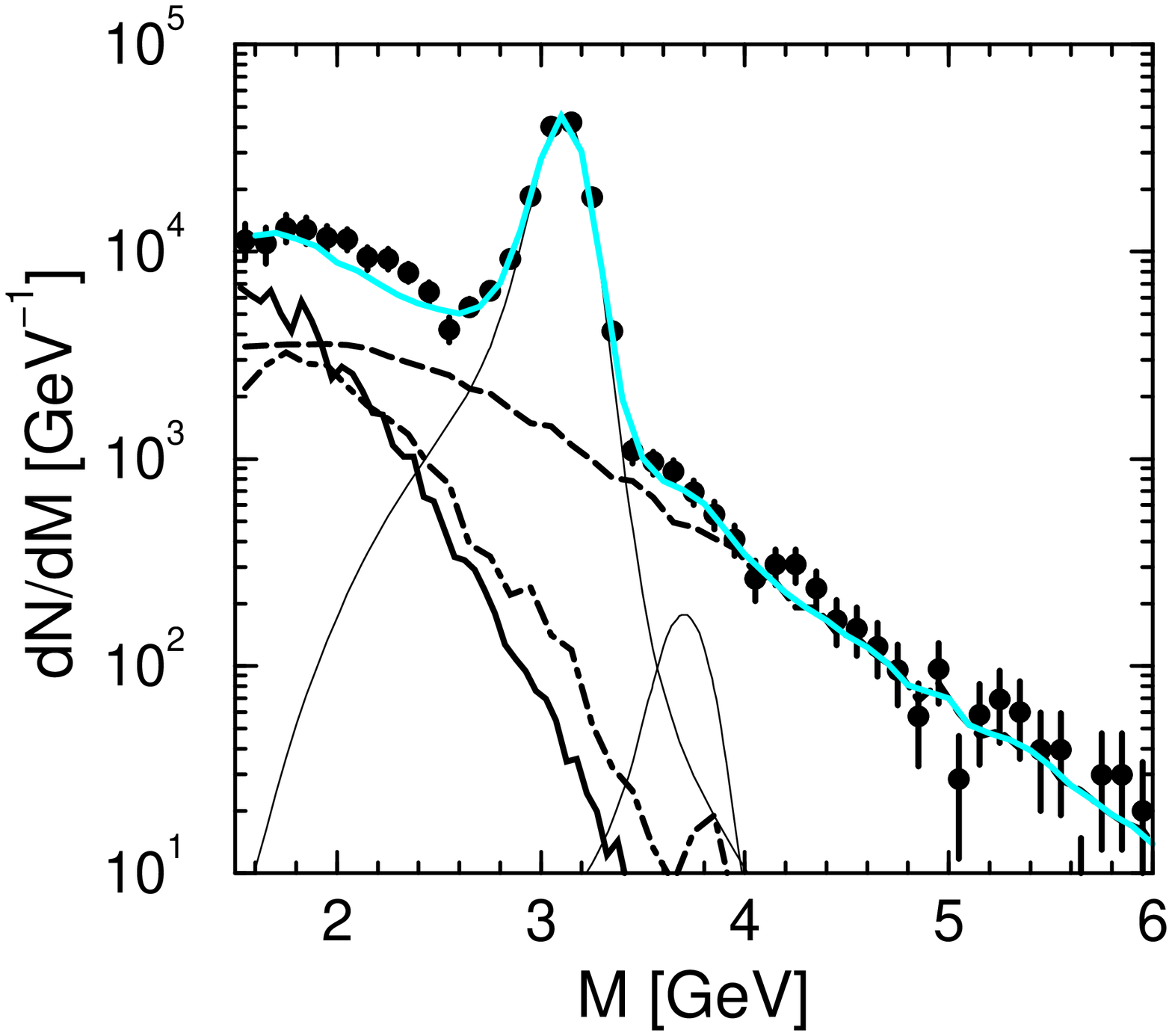,width=6cm,angle=-90}
\hfill
\psfig{file=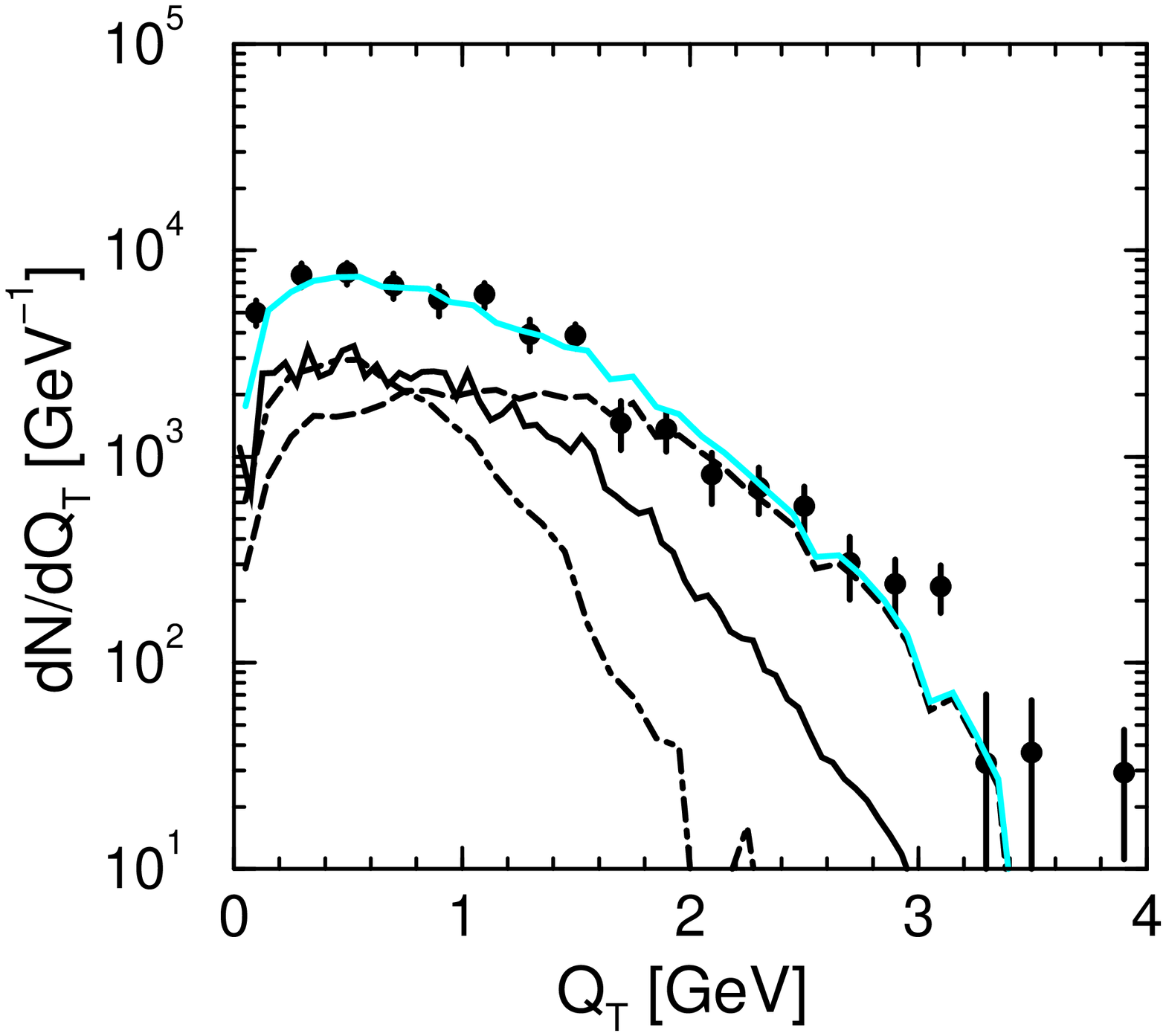,width=6cm,angle=-90}
~\\[.5cm]
\caption{
Comparison of our calculations with NA50 data \cite{NA50}.
Left panel: invariant mass distribution,
right panel: $Q_\perp$ spectrum for the invariant
mass bin $M = 2.1 \cdots 2.7$ GeV
(solid curves: thermal contribution,
dot-dashed curves: open charm,
dashed curves: Drell-Yan,
thin lines: parametrizations of the $J/\psi$ and $\psi'$
contributions according to \cite{NA50},
gray curves: sum of these contributions).
}
\label{f_2}
\end{figure}

The resulting invariant mass and transverse momentum spectra,
including the thermal source contribution, are displayed in Fig.~\ref{f_2}.
The thermal source, with parameters $T_{\rm eff}$ and $N_{\rm eff}$
adjusted to the above CERES data,
is needed to achieve the overall agreement with data.
This unifying interpretation of different measurements has to be
contrasted with other proposals of explaining the dimuon excess in the
intermediate mass region either by final state hadron interactions
\cite{Lin_Wang}
or by an abnormally large open charm enhancement \cite{NA50}. 
The latter one should
be checked experimentally \cite{NA6i} thus attempting a firm
understanding of dilepton sources.
Notice that our minimum parameter model describes the data equally well
as the more detailed dynamical models \cite{RappShuryak,Phys.Lett.}.

\subsection{Sulfur beam data} 

\begin{figure}[t] 
\centering
~\\[-.1cm]
\psfig{file=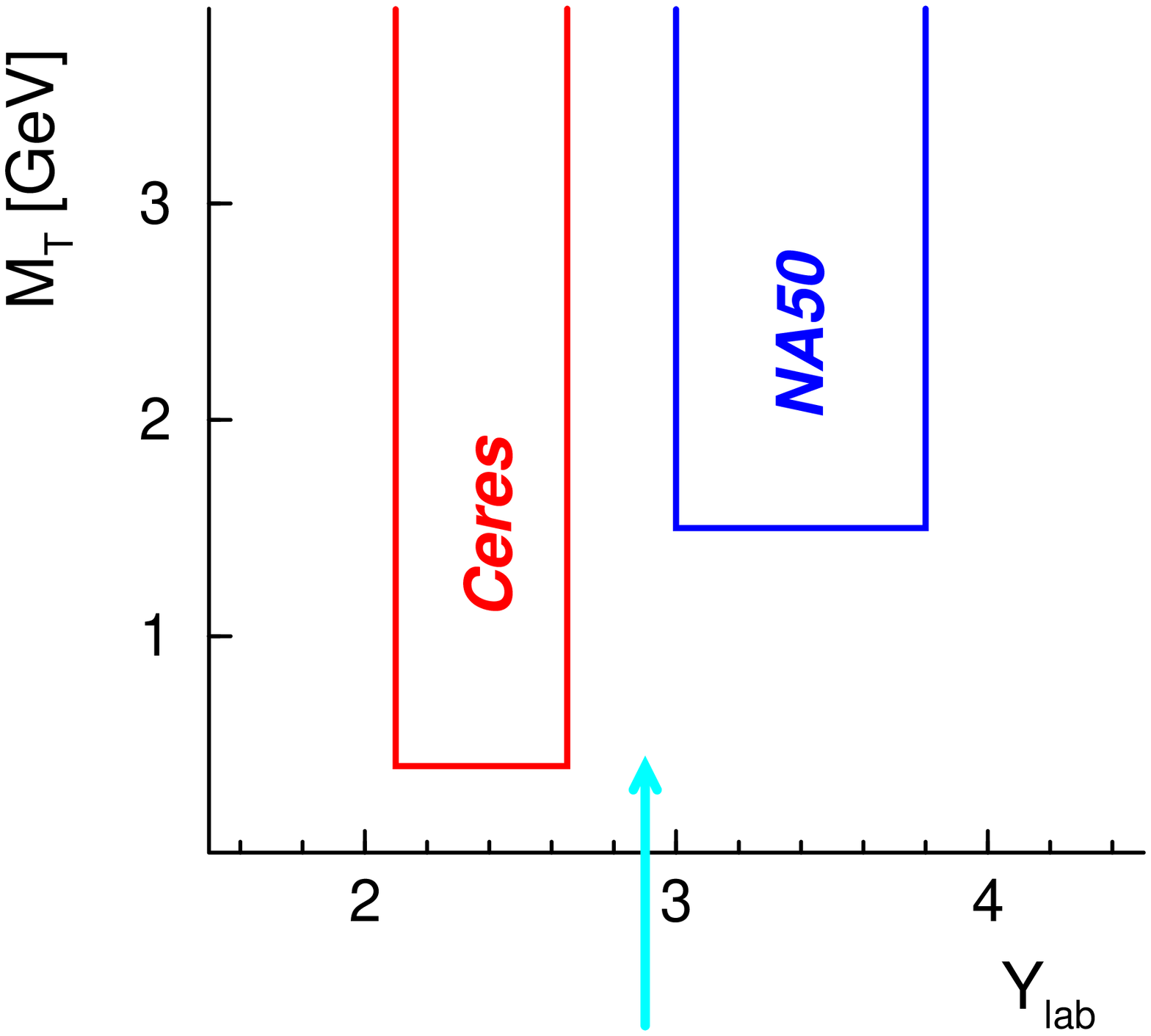,width=7cm,angle=-90}
\hfill
\psfig{file=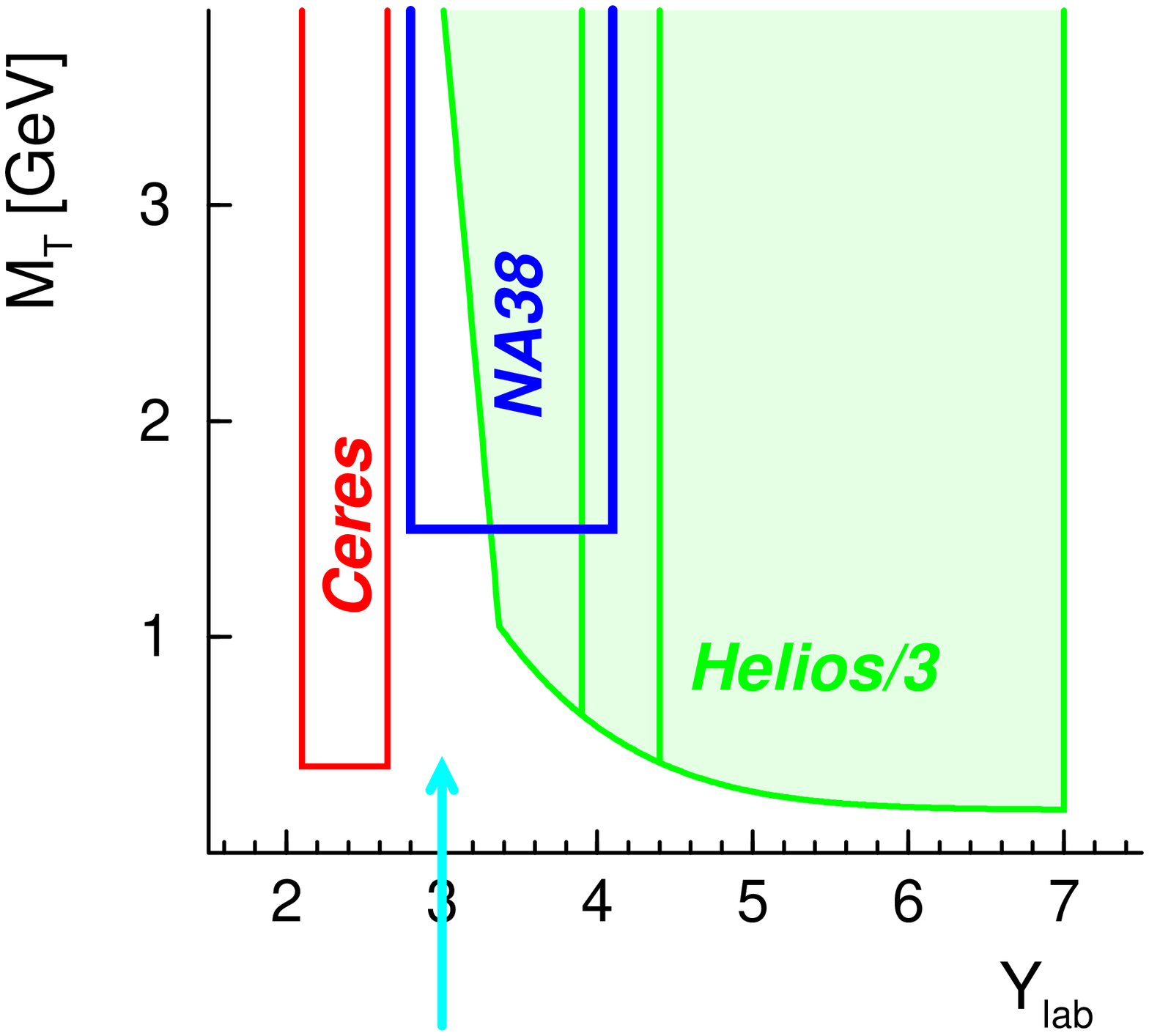,width=7cm,angle=-90}
~\\[.5cm]
\caption{
Coverage of the rapidity $Y_{\rm lab}$ and transverse mass $M_\perp$ of the
various dilepton experiments.
(a) left panel: lead beam,
(b) right panel: sulfur beam.
The arrows indicate the $pp$ center-of-mass rapidity. 
}
\label{f_3}
\end{figure}

\begin{figure}[t] 
\centering
~\\[-.1cm]
\psfig{file=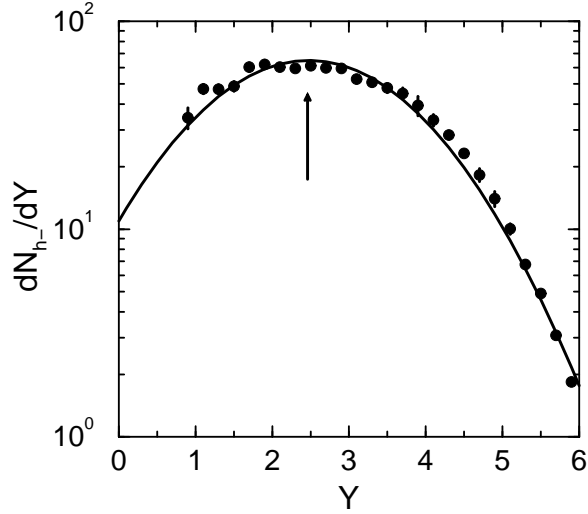,width=7cm,angle=-90}
~\\[.5cm]
\caption{
A Gaussian fit to the NA35 data of negatively charged hadrons
\cite{NA35}. 
The arrow indicates
the center of the Gaussian which has a width of $\sigma_{h^-} = 1.1$. 
}
\label{f_15}
\end{figure}

Let us now turn to the older sulfur beam data (cf.\ 
\cite{RappWambach}
for a recent survey on models and an extensive reference list). 
Since a much larger rapidity interval is covered
(see Fig.~\ref{f_3})
we smear the source distribution (3) by a Gaussian function with a width
of $\sigma = 0.8$, i.e.\ 
\begin{equation}
\frac{dN}{d^4 Q} = 
\frac{5 \alpha^2}{36 \pi^4} N_{\rm eff}
\int \frac{dY'}{\sqrt{2 \pi \sigma^2}} 
\exp \left\{ - 
\frac{M_\perp \cosh (Y - Y')}{T_{\rm eff}} \right\}
\exp \left\{ \frac{(Y' - Y_{\rm cms})^2}{2 \sigma^2} \right\}.
\end{equation}
For $Y_{\rm cms}$ we choose 2.45 
as suggested by an analysis of the
rapidity distribution of negatively charged hadrons
from NA35 \cite{NA35} as displayed in Fig.~\ref{f_15}.
The width of the dilepton source, $\sigma$, 
is somewhat smaller than the width of the hadron distribution,
cf.\ Fig.~\ref{f_15}. 
Here we neglect also the small differences of the various target nuclei
(Au, U, W)
and attempt a unique parametrization of the $e^- e^+$ and 
$\mu^- \mu^+$ channels.
 
\subsubsection{CERES experiment \label{CERES_sulfur}} 

\begin{figure}[t] 
\centering
~\\[-.1cm]
\psfig{file=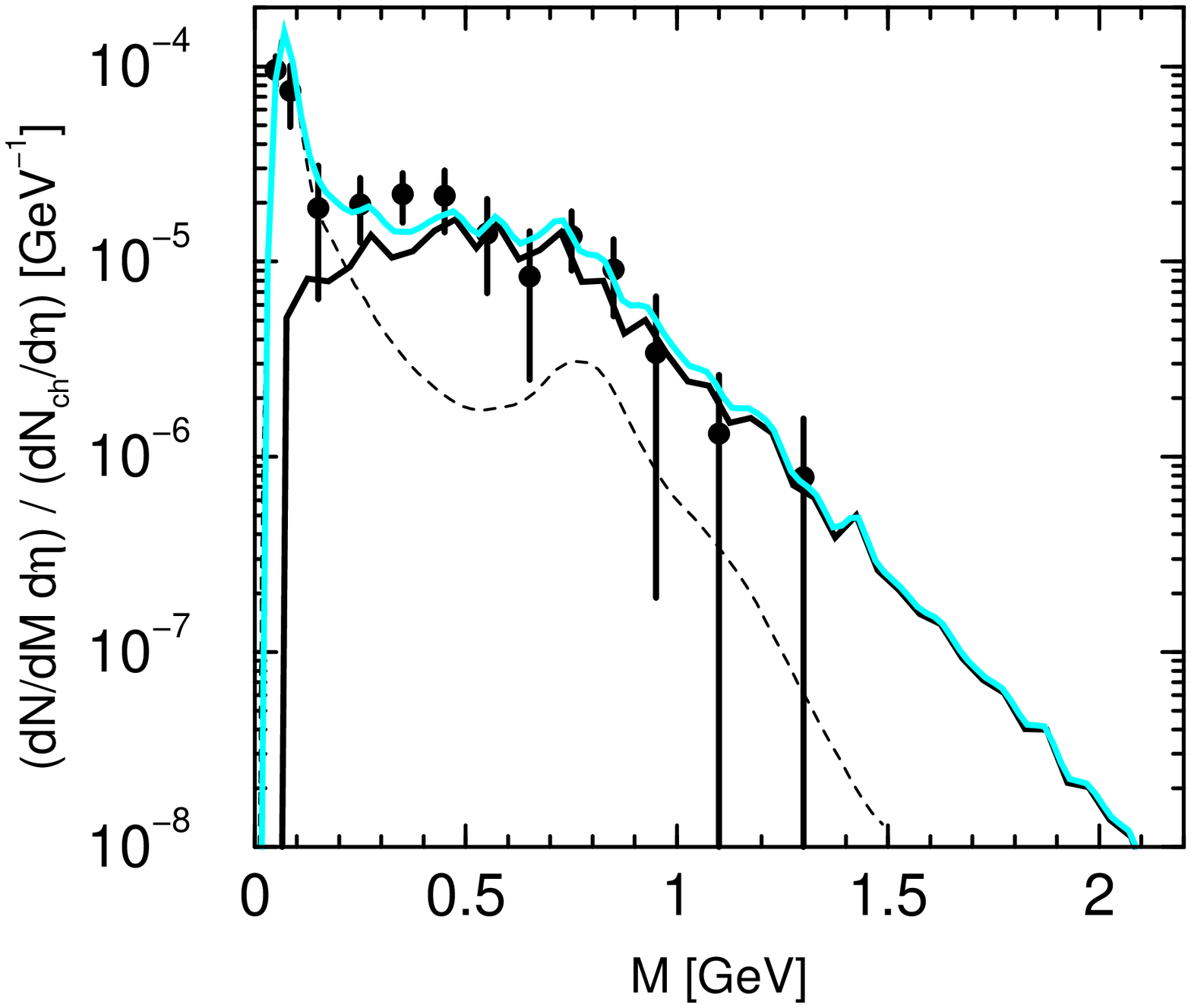,width=6cm,angle=-90}
\hfill
\psfig{file=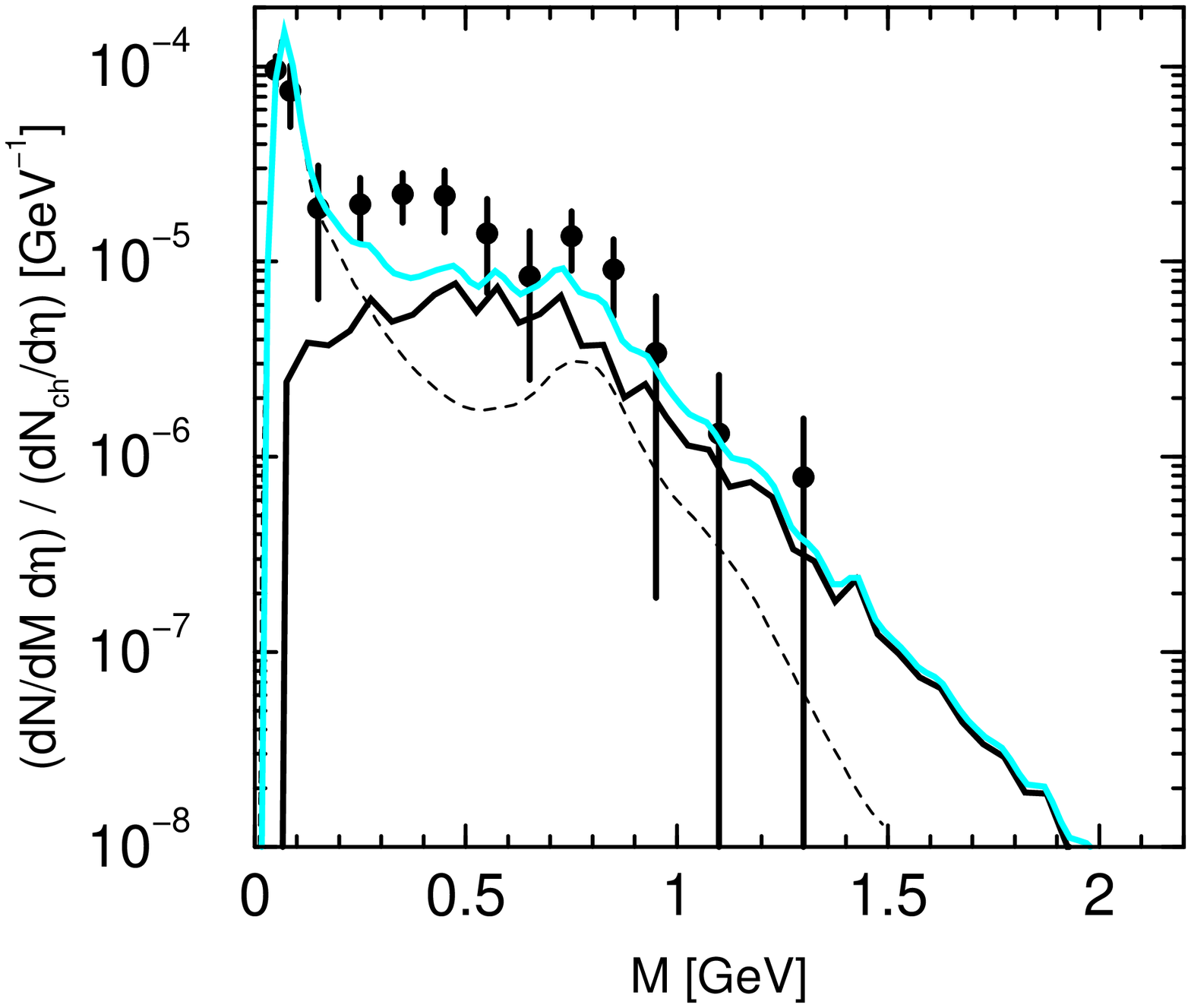,width=6cm,angle=-90}
~\\[.5cm]
\caption{
Comparison of our model calculations with the CERES data
\cite{CERES_S} with separately adjusted normalization factor
$N_{\rm eff} = 11.2 \times 10^4$ fm${}^4$
(left panel) or with normalization 
$N_{\rm eff} = 5.3 \times 10^4 $ fm${}^4$
adjusted to HELIOS/3 data \cite{HELIOS-3} (right panel). 
Dashed curves: hadronic cocktail from \cite{CERES_S},
solid curves: thermal yield,
gray curves: sum.
}
\label{f_5}
\end{figure}

For the CERES $e^+ e^-$ data \cite{CERES_S} in central 
S(200 AGeV) + Au reactions, corresponding to 
$\langle \frac{dN_{ch}}{d \eta} \rangle = 125$,
the published hadronic cocktail is used.
The acceptance is the same as described in subsection \ref{CERES_lead}.
The comparison of our calculations with the data is displayed in 
Fig.~\ref{f_5}. A good description of the data is achieved
by $T_{\rm eff} = 160$ MeV. One observes in the left pannel of
Fig.~\ref{f_5} indeed a fairly well 
reproduction of the spectral shapes for the choice of the
normalization factor $N_{\rm eff} = 11.2 \times 10^4$ fm${}^4$.
(Notice that this normalization factor is ununderstandably large.
We focus here, however, on the shape of the spectra
and do not attempt a change of our simple parametrization to resolve this
issue, e.g.\ by employing another rapidity distribution.
In this context we mention the second reference in
\cite{V.Koch} where, within the same transport code, the lead beam
data \cite{CERES_Pb} are satisfactorily reproduced but the sulfur
beam data \cite{CERES_S} strongly underestimated.
Similarily, if we use the same normalization factor, as adjusted to
the HELIOS/3 data (see subsection 3.2.3 below) our resulting spectrum
is below the data and even outside of the systematical error bars
at $M \sim 400$ MeV, see right pannel of Fig.~\ref{f_5}.)

\begin{figure}[t] 
\centering
~\\[-.1cm]
\psfig{file=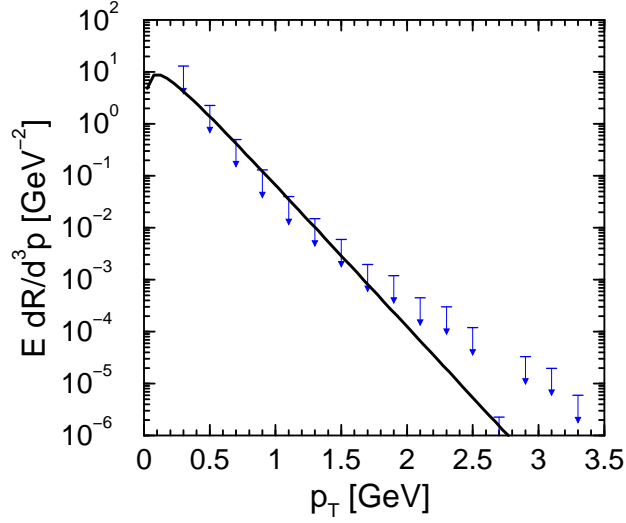,width=7cm,angle=-90}
~\\[.5cm]
\caption{
A comparison of the thermal photon spectrum 
(without transverse matter flow)
with the experimental
upper bounds \cite{WA80} when adjusting the source strength to the 
HELIOS/3 data
\cite{HELIOS-3}.
}
\label{f_9}
\end{figure}

We mention additionally that adjusting the normalization to the HELIOS/3 data
\cite{HELIOS-3} 
the published upper bounds of the direct photon yields \cite{WA80} 
are just compatible with our
model calculations when adopting the model described in section~2 
for photons \cite{our_photons}
(see Fig.~\ref{f_9}). In contrast, when choosing the
normalization, which delivers an optimum description of the CERES
data, one would be above the upper photon bounds in \cite{WA80}.

\subsubsection{NA38 experiment} 

\begin{figure}[t] 
\centering
~\\[-.1cm]
\psfig{file=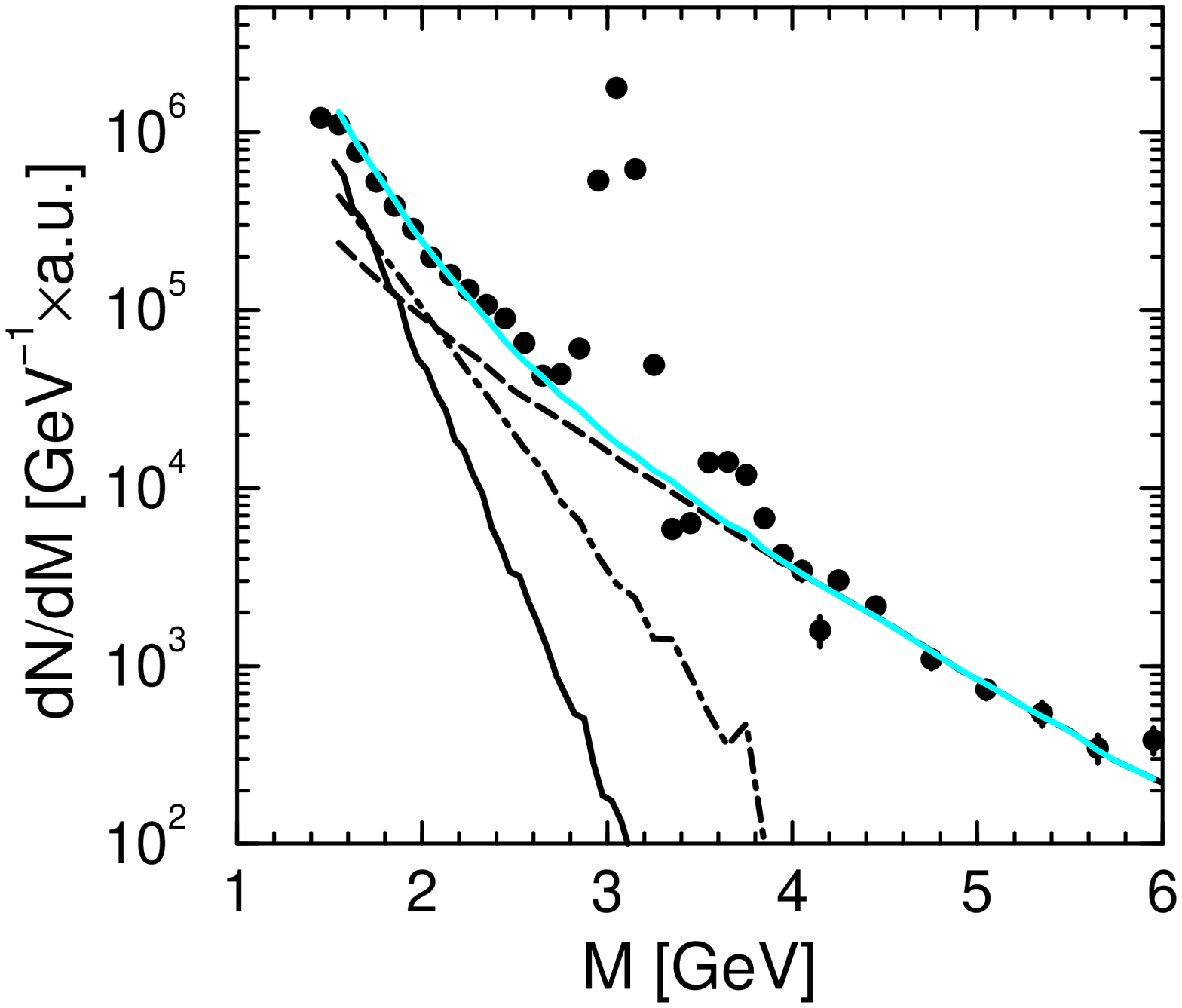,width=6cm,angle=-90}
\hfill
\psfig{file=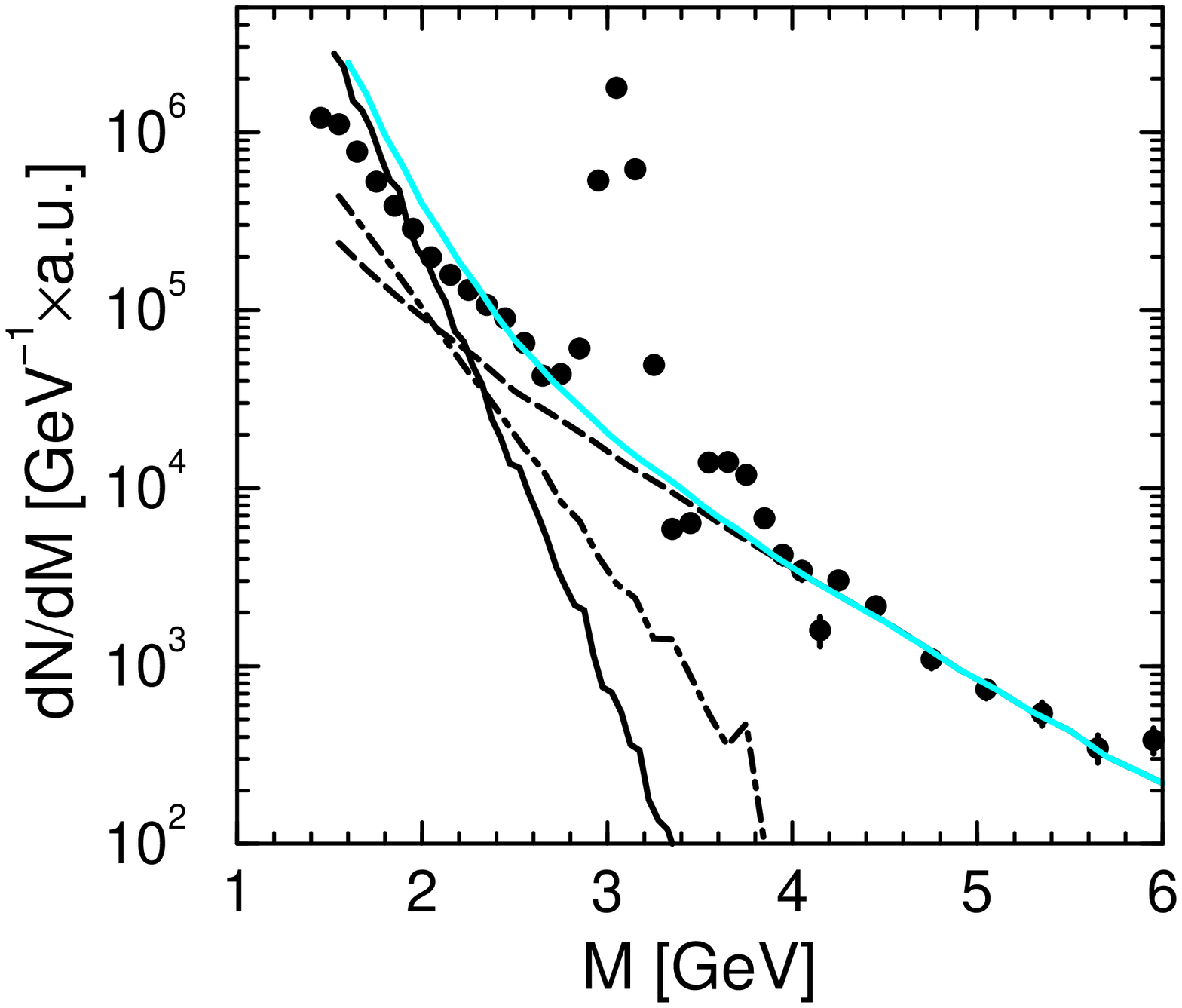,width=6cm,angle=-90}
~\\[.5cm]
\caption{
Comparison of our model calculations with the NA38 data
\cite{NA38} with separately adjusted normalization factor
$N_{\rm eff} = 21.3 \times 10^4$ fm${}^4$ (left panel)
and with normalization 
$N_{\rm eff} =  5.3 \times 10^4$ fm${}^4$
adjusted to HELIOS/3 data \cite{HELIOS-3} (right panel).
Meaning of the curves: thermal yield, open charm contribution and Drell-Yan
(from left to right at larger $M$);
gray curves: sum of all contributions;
}
\label{f_6}
\end{figure}

The acceptance of this dimuon experiment is described by
$Y_{\rm lab} = 2.8 \cdots 4.1$ and
$\vert \cos \Theta_{\rm CS} \vert < 0.5$.
The comparison of our calculations with the NA38 data \cite{NA38}
for the reaction S(200 AGeV) + U
in the intermediate-mass and high-mass region is displayed in
Fig.~\ref{f_6}. Adjusting the normalization factor to the data
we achieve an optimum description by 
$N_{\rm eff} = 1.3 \times 10^4$ fm${}^4$ (left pannel),
while the use of the normalization, adjusted to HELIOS/3 data,
results in a unsatisfactory data description
(right pannel in Fig.~\ref{f_6}). 
It should be emphasized that all other data sets we are analyzing
are for more restrictive central events. It is therefore clear that
the required normalization factor for the NA38 data is smaller.

\begin{figure}[t] 
\centering
~\\[-.1cm]
\psfig{file= 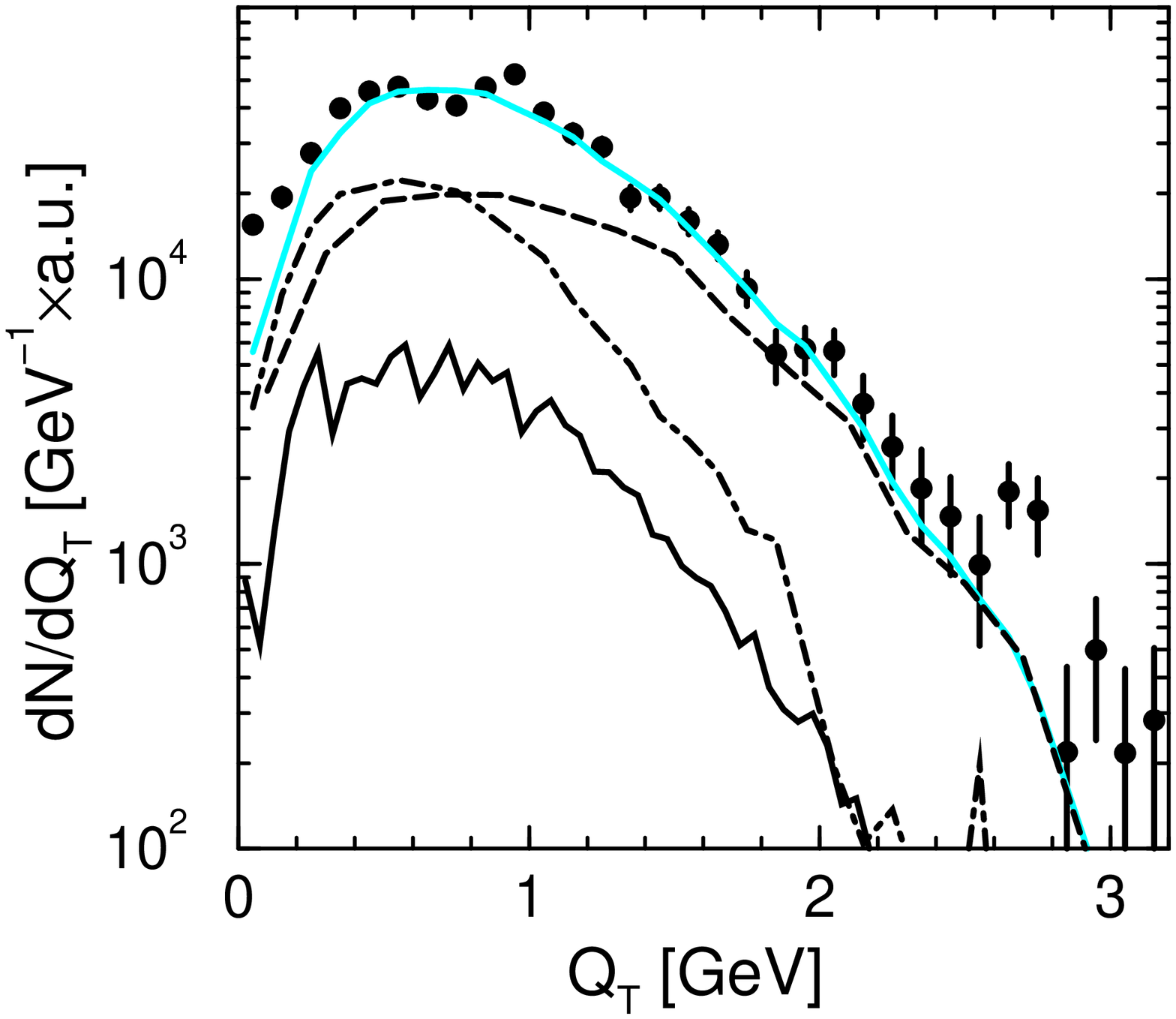,width=6cm,angle=-90}
\hfill
\psfig{file= 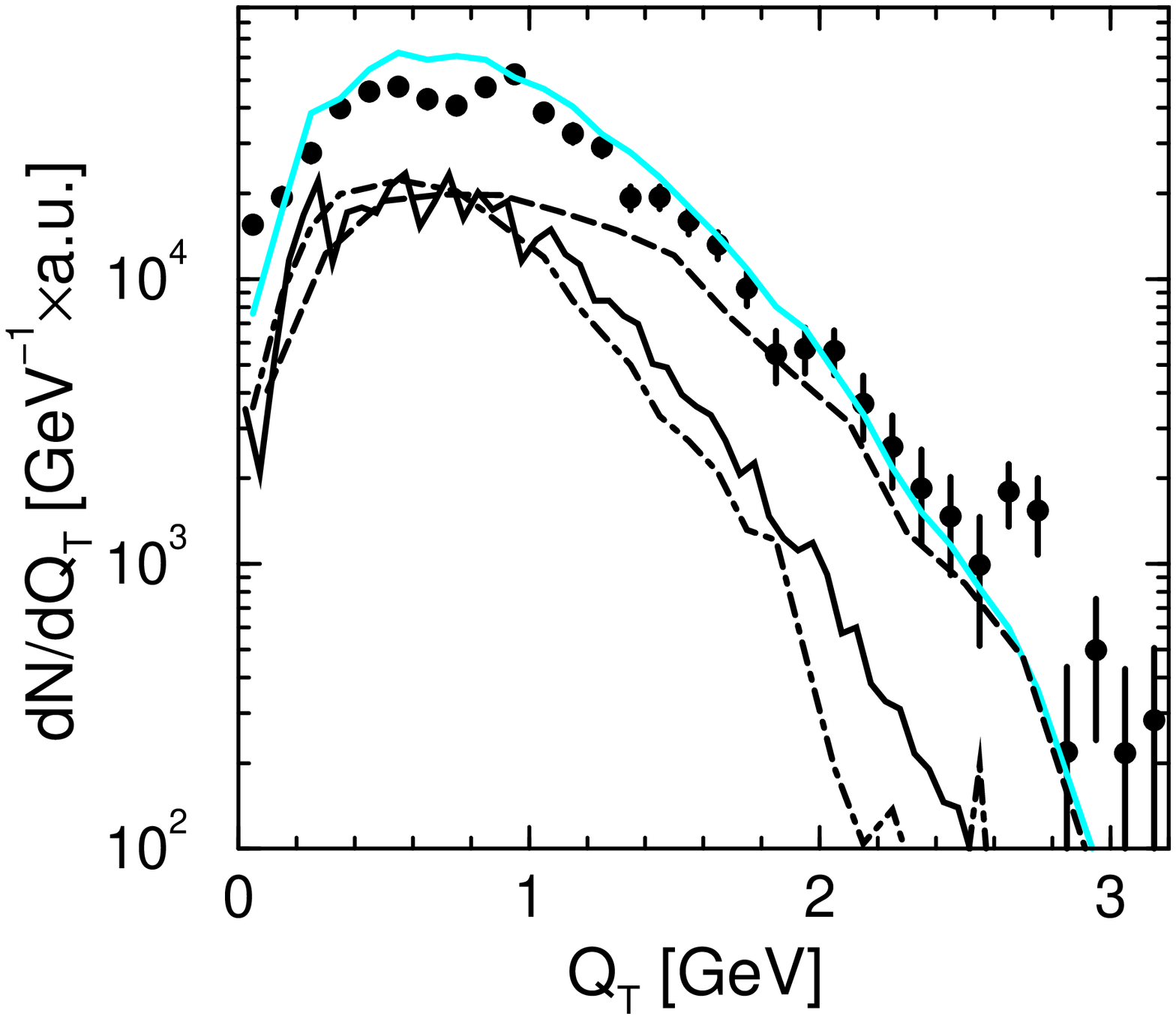,width=6cm,angle=-90}
~\\[.5cm]
\caption{
Comparison of our model calculations with the transverse momentum spectrum of
dileptons from NA38 \cite{NA38} in the intermediate-mass region
$M = 2.1 \cdots 2.7$ GeV.
Normalizations as in Fig.~\protect\ref{f_6}.
}
\label{f_8}
\end{figure}

The available transverse momentum spectrum of NA38 is also nicely reproduced
in shape (see Fig.~\ref{f_8}). 

Recently, the NA38/50 collaboration published also dilepton data in the
low-mass region \cite{NA38/50}. 
Since we have no reliable background contribution
(hadronic cocktail) within the given acceptance 
at hand we discard the inclusion of these data in our analysis. 

\subsubsection{HELIOS/3 experiment} 

While we can nicely reproduce the Drell-Yan background for the 
HELIOS/3 experiment (cf.\ \cite{HELIOS-3}),
our PYTHIA simulations deliver another open charm contribution 
than the one used
in previous analyses \cite{LiGale}. 
Since the accurate knowledge of the background
contributions is necessary prerequisite, we use therefore for our analysis the
difference $\mu^+ \mu^-$ spectra of S(200 AGeV) + W and p(200 GeV) + W 
reactions 
\cite{HELIOS-3},
thus hoping to get rid of the background since these 
spectra are appropriately normalized. 
The centrality selection for these data is a multiplicity larger than
100 in the pseudo-rapidity bin $3.5 \cdots 5.2$ resulting in an
averaged multiplicity of 134.6 \cite{HELIOS-3}.

The acceptance is described by 
$M_\perp > max \left\{4 ( 7 - 2 Y_{\rm lab}) \mbox{GeV},
\sqrt{4 m_\mu^2 + 
\left( \frac{15 {\mbox{{\small GeV}}}}{\cosh Y_{\rm lab}} 
\right)^2} \right\}$
and the data are binned in the rapidity intervals
$Y_{\rm lab} = 3.0 \cdots 3.9, 3.9 \cdots 4.4, 4.4 \cdots 7.0$.
It is in particular this wide rapidity span and the binning which
require the Gaussian smearing of the dilepton source.
Considering the rapidity-integrated yield, as done in most
previous analyses when only the integrated data were at disposal, 
discards an important information
and allows a unique description of the CERES lead beam data and HELIOS/3 
sulfur beam data without any problem \cite{Cassing_2}.

\begin{figure}[t] 
\centering
~\\[-.1cm]
\psfig{file=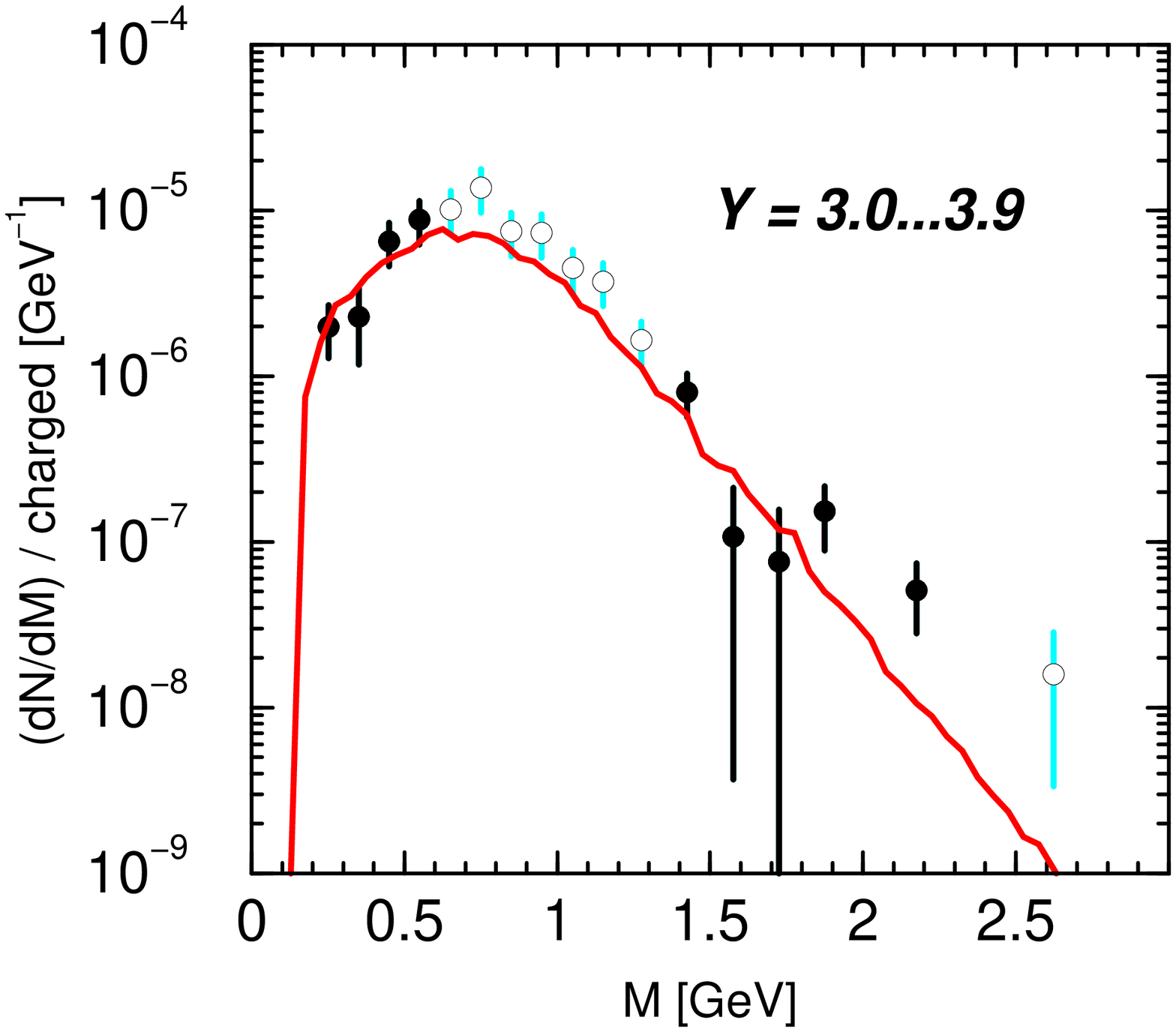,width=5.1cm,angle=-90}
\hfill
\psfig{file=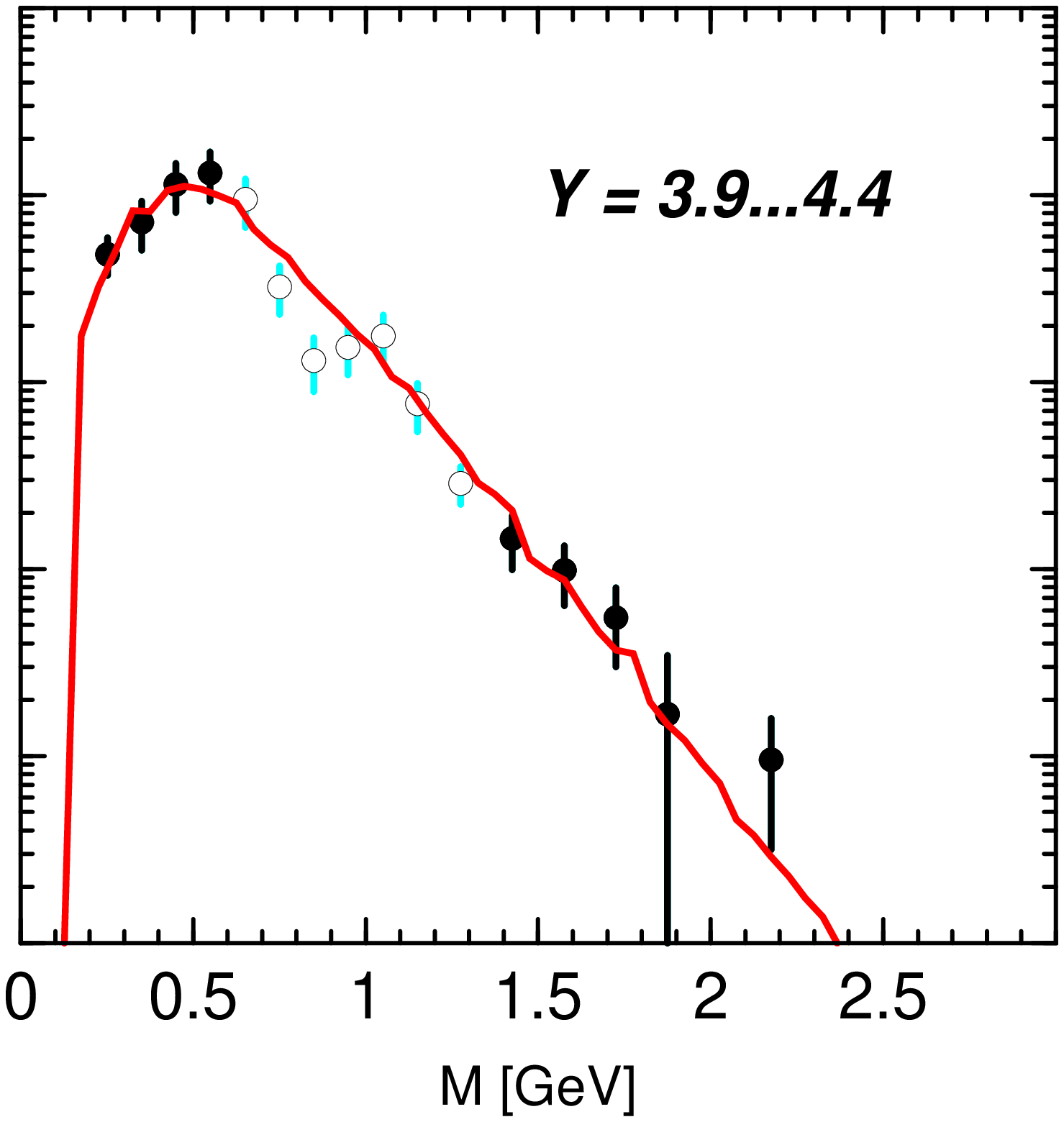,width=5.1cm,angle=-90}
\hfill
\psfig{file=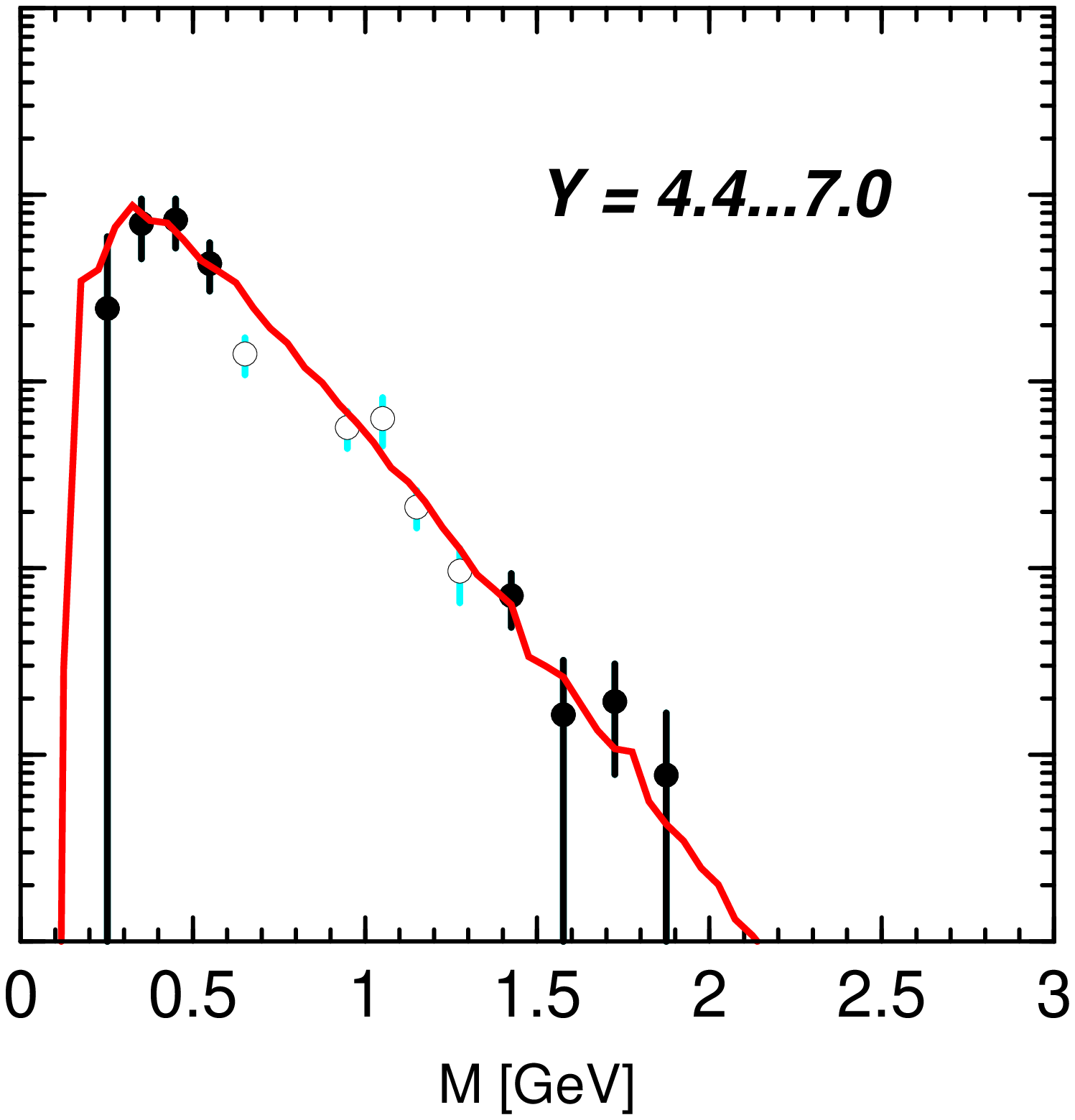,width=5.1cm,angle=-90}
~\\[.5cm]
\caption{
Comparison of our model calculations with the normalized 
HELIOS/3 difference data
[S(200 A$\cdot$GeV) + W minus p(200 A$\cdot$GeV) + W]
in various rapidity bins
\cite{HELIOS-3} with a unique normalization factor
$N_{\rm eff} = 5.3 \times 10^4$ fm${}^4$.
The black dots indicate data groups for which also $Q_\perp$ spectra
are available (cf.\ Fig.~\protect\ref{f_7_1}).
}
\label{f_7}
\end{figure}

\begin{figure}[t] 
\centering
~\\[-.1cm]
\psfig{file=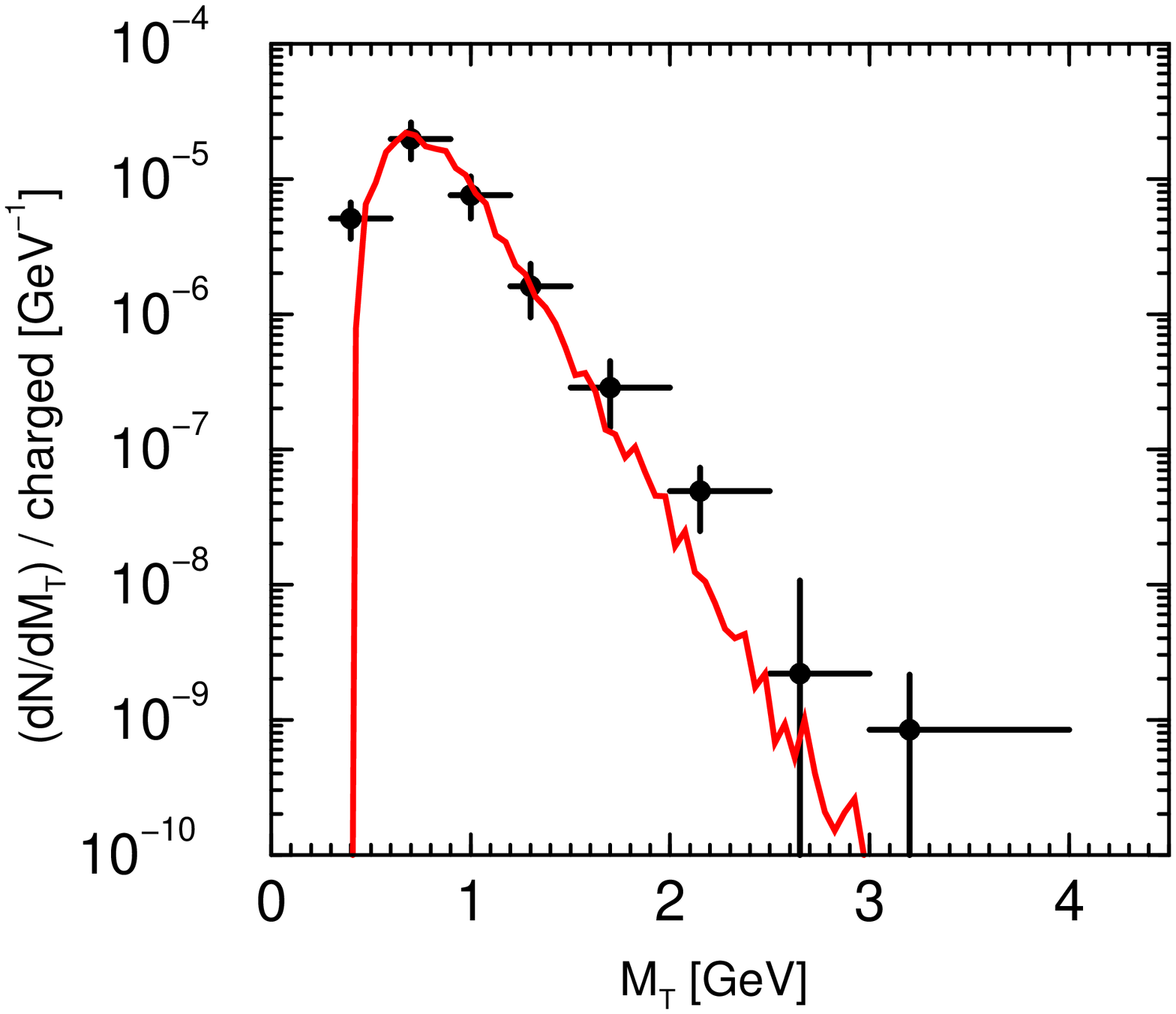,width=5.9cm,angle=-90}
\hfill
\psfig{file=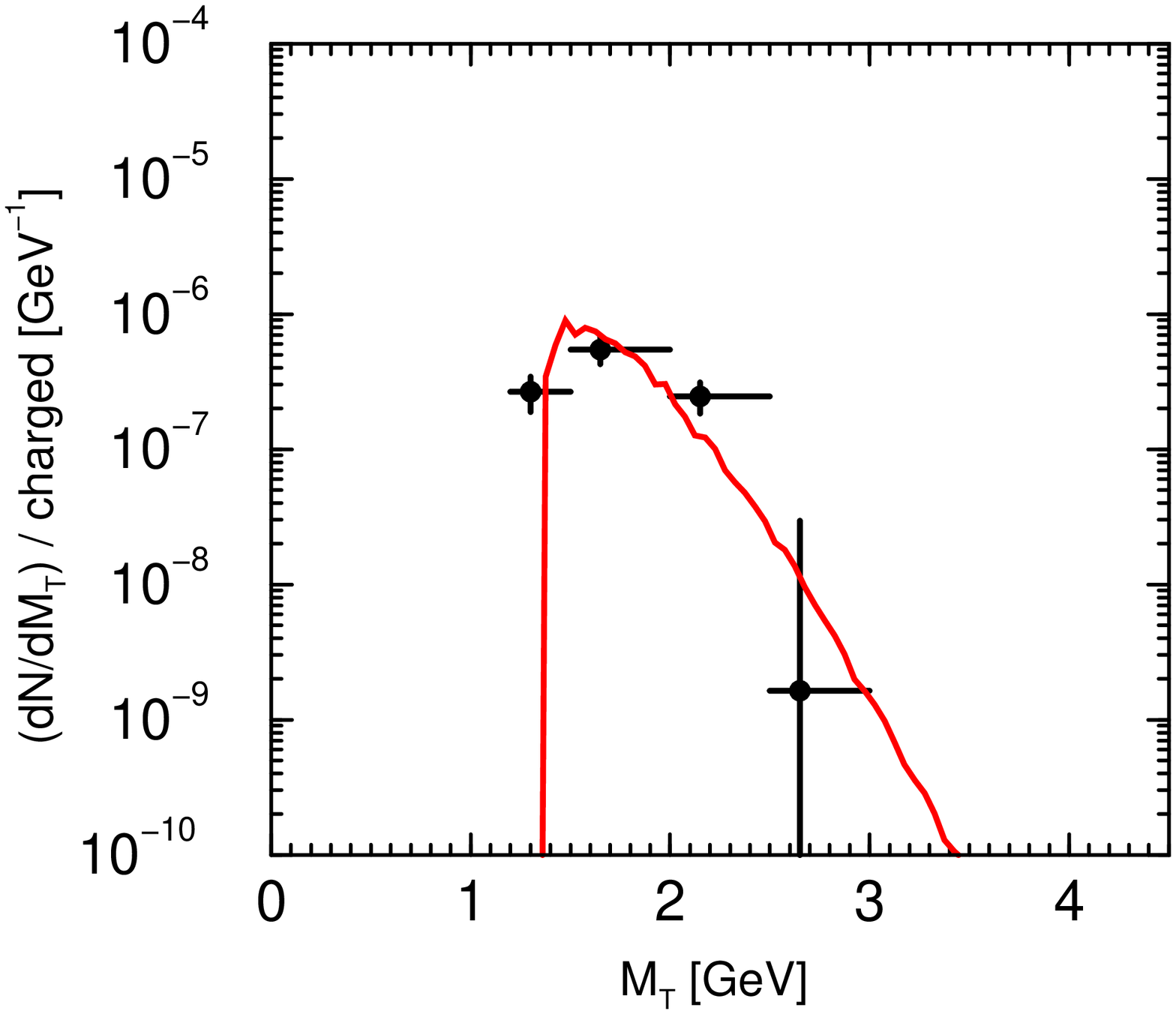,width=5.9cm,angle=-90}
~\\[.5cm]
\caption{
Comparison of our model calculations with the HELIOS/3 difference data
\cite{HELIOS-3} for the $M_\perp$ distributions
with a unique normalization factor
$N_{\rm eff} = 2 \times 5.3 \times 10^4$ fm${}^4$
within the rapidity bin $Y_{\rm lab} = 3.0 \cdots 4.4$
for the invariant mass intervals $M = 0.2 \cdots 0.6$ GeV
(left panel)
and $M = 1.35 \cdots 2.5$ GeV
(right panel).
The additional factor 2 in the normalization is from a mismatch
of these data compared with the ones in Fig.~\protect\ref{f_7}.
}
\label{f_7_1}
\end{figure}

As seen in Fig.~\ref{f_7} the difference spectra in the various
rapidity bins
are fairly well described for $N_{\rm eff} = 5.3 \times 10^4$
fm${}^4$. (Notice that, in comparison with the lead beam data,
also this normalization is quite large.)
The $M_\perp$ spectra in the two available $M$ bins for the
integrated rapidity bin $Y = 3 \cdots 4.4$ are also
described in gross features, see Fig.~\ref{f_7_1}. 
(Integrating the experimental $M_\perp$ spectra in this figure
one finds a factor of 2 difference to the spectra in Fig.~\ref{f_7}
when integrating these over the corresponding $M$ intervals.
This is accounted for in the Fig.~\ref{f_7_1}.) 

\section{Discussion of flow and time evolution effects} 

\subsection{Flow effects}

\begin{figure}[t] 
\centering
~\\[-.1cm]
\psfig{file=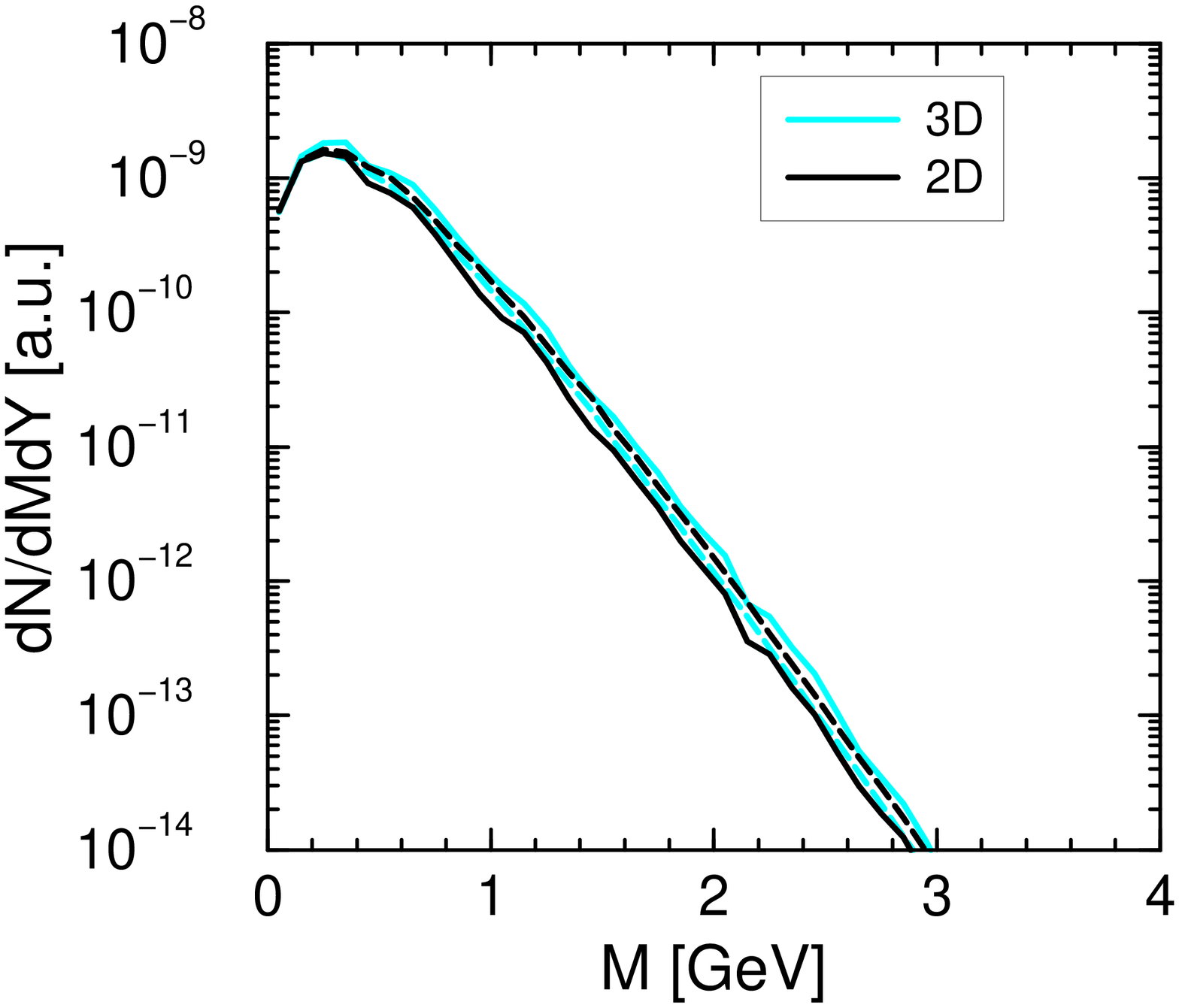,width=6cm,angle=-90}
\hfill
\psfig{file=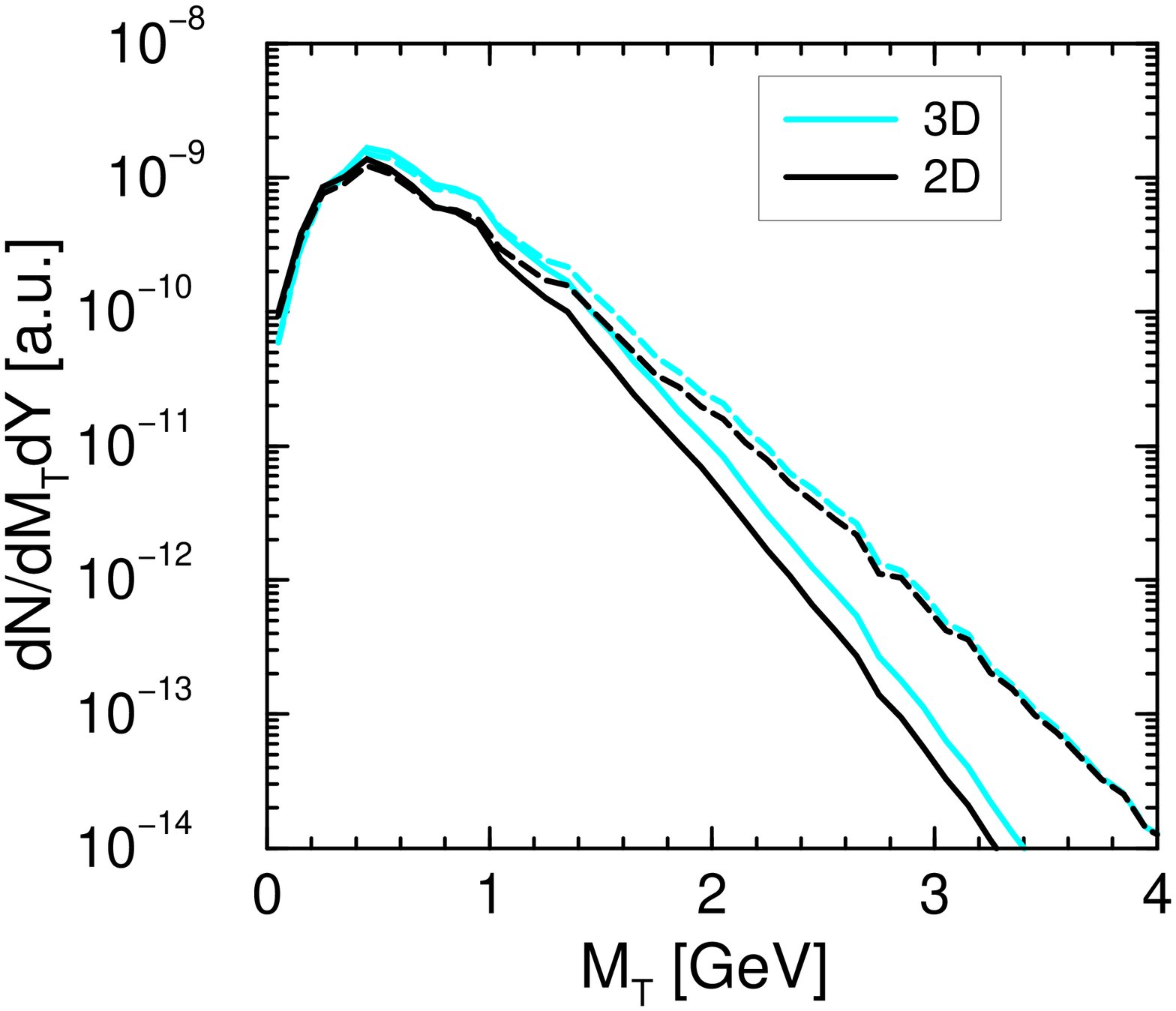,width=6cm,angle=-90} 
~\\[.5cm]
\caption{
Sensitivity of the dilepton spectra at midrapidity against transverse
flow within the blast wave model.
Left panel: $d N/d M \, d Y$,
right panel: $d N/d M_\perp \, d Y$.
Solid curves: without transverse flow,
dashed curves: with transverse flow $v_\perp =$ 0.6 (left panel) or
0.4 (right panel).
The black (gray) curves are for a 2 (3) dimensional expansion
(cf.\ \protect\cite{Phys.Lett.} for further details).
}
\label{f_11}
\end{figure}

As qualitatively discussed in \cite{Phys.Lett.} flow effects affect
mainly the $Q_\perp$ or $M_\perp$ spectra.
Explicit formulae for the dilepton spectra are given in
\cite{Phys.Lett.} for 
(i) spherical expansion and
(ii) longitudinally boost-invariant expansion superimposed on transverse
expansion. In Fig.~\ref{f_11} we show the result of Monte Carlo simulations
of the rates $dN /d M \, d Y$ and $dN/dM_\perp \, d Y$ 
at $Y = 0$ for
various values of the transverse flow velocity. One observes
that the invariant mass spectra,
delivered by a Monte Carlo procedure for generating the distribution
Eq.~(4), are indeed insensitive against
flow (see Fig.~\ref{f_11}, left panel).
This is known for case (ii) for some time \cite{Kajantie}.
In contrast, the $Q_\perp$ spectra are sensitive against flow
as demonstrated in the right panel of  Fig.~\ref{f_11},
in particular in the large-$Q_\perp$ region. 
However, as seen in Figs.~\ref{f_1} and \ref{f_2}, 
the large $Q_\perp$ region is dominated
by the background contributions (hadronic cocktail or Drell-Yan).
Therefore, one can indeed neglect the flow. Note that the key for this
statement is a constraint on transverse flow from hadronic data. The analyzes
of the $m_\perp$ spectra of several hadron species point to flow velocities
$\langle v_\perp \rangle$ in the range from 0.43 c \cite{J.Phys., Cassing} 
to 0.55 c \cite{Heinz}.
Since the flow is expected to increase continuously with time due to
acceleration from pressure, the transverse flow
of hadrons at kinetic freeze-out is a temporal maximum value.
The spatial and timewise average of the flow, relevant for dilepton
and photon emission, must be smaller than the quoted values. 

The situation may change at RHIC and LHC,
where higher initial temperatures, and consequently larger pressures,
cause a stronger transverse flow which could be manifest in dilepton
$Q_\perp$ spectra.

\subsection{Time evolution effects}

\begin{figure}[t] 
\centering
~\\[-.1cm]
\psfig{file=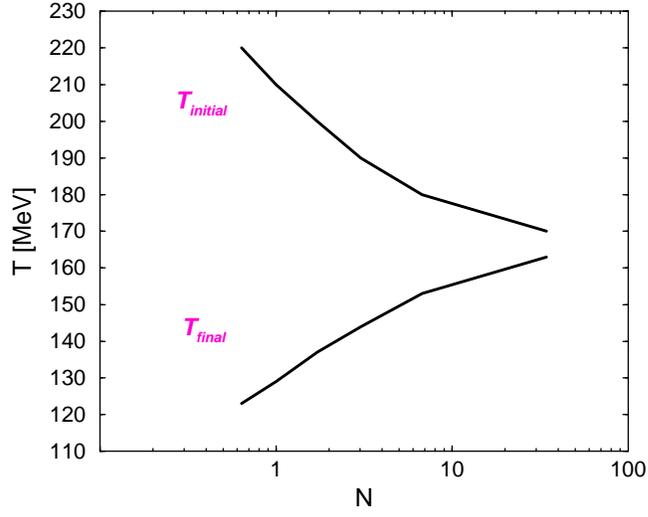,width=7cm,angle=-90}
~\\[.5cm]
\caption{
The change of the initial and final temperatures as a function
of the normalization factor $N$.
}
\label{f_16}
\end{figure}

Let us now discuss time evolution effects.
As an example we show in Fig.~\ref{f_16} such initial temperatures
and the corresponding final temperatures as a function of an 
normalization factor, $N$, 
which deliver the same invariant mass spectra in the range
covered by the CERES and NA50 lead beam experiments displayed in
Figs.~\ref{f_1} and \ref{f_2}. 
In calculating the curves in Fig.~\ref{f_16}
the time evolution according to
$T = T_\infty + (T_{\rm initial} - T_\infty) 
\exp \left\{ - \frac{t}{t_2}\right\}$,
$V(t) = N \frac{A + B}{2.5 n_0} \exp \left\{ \frac{t}{t_1}\right\}$,
with 
$T_\infty = 110$ MeV, $t_1 = 5$ fm/c, $t_2 = 8$ fm/c
is used as in \cite{Phys.Lett.}
(cf.\ \cite{KlinglWeise}), 
where $A, B$ are the mass numbers of the
colliding nuclei, $n_0 = 0.17$ fm${}^{-3}$.
The temperature evolution starts at $T_{\rm initial}$ and is stopped
at $T_{\rm final}$.
(Such a temperature and volume evolution have been used first
in \cite{Klingl_Weise} to show that the superposition of the 
thermal $q \bar q$ annihilation rate and the hadronic cocktail
reproduces the shape of the CERES data obtained with the sulfur beam 
\cite{CERES_S}.)
From Fig.~\ref{f_16} one can infer that, if the final temperature 
$T_{\rm final}$ is
identified with the hadron kinetic freeze-out temperature of, e.g.,
125 MeV according to \cite{J.Phys.}, the corresponding initial temperature
$T_{\rm initial}$
would be 215 MeV. This statement, however, depends on the
assumed temperature evolution. For instance, in case of a bag model 
equation of state a different weighting would occur and therefore
different values of the initial and final temperatures would follow.
Nevertheless, the merging of the initial and final temperatures,
$T_{\rm initial}$ and $T_{\rm final}$, 
at 170 MeV, displayed in Fig.~\ref{f_16}, suggests to use only
$T_{\rm eff}$ and $N_{\rm eff}$ instead of more parameters.
Indeed, the above time evolution equations and the values displayed in
Fig.~\ref{f_16} reproduce the value of $N_{\rm eff} = \int dt \, V(t)$
as used in subsection 3.1.

\section{Summary} 

An attempt is reported to explain nearly the whole set of
the dilepton data in recent CERN-SPS
experiments with heavy-ion beams. 
We assume that the dilepton emissivity is the same as that in 
$q \bar q$ annihilation.
This hypothesis is made to work in the low mass region and in the
intermediate mass region possibly for different reasons. Whether
several channels add up to smooth out a spectrum with intrinsic
structure (``kinematical saturation'') or whether those structure are
broadened and washed out is not addressed here. 
Note that 
while we find a good overall reproduction of the shapes of the experimental
spectra, only the lead beam data can be explained with a unique and reasonable
normalization.

As main result we emphasize that the average temperature of 
$T_{\rm eff} = 160 \cdots 170$ MeV used in this study  
is in good agreement with the
temperature deduced from measured hadron ratios \cite{abundances}.
Since $T_{\rm eff}$ is an average, one can conclude that the achieved
maximum temperatures are above this value and, therefore, in a region
where deconfinement is expected according to lattice QCD
calculations \cite{lattice_QCD}, which presently advocate
a deconfinement temperature of $T_c = 170$ MeV or somewhat larger. 

In the present work we restrict ourselves to central collisions
(except for the NA38 data)
and compare several experiments with $e^+ e^-$ and $\mu^+ \mu^-$
channels.
On the basis of this study  an analysis of the $E_\perp$ dependence of the 
combined NA38 and NA50 data seems feasible. 
This is of interest since, due to the impact parameter dependence,
some interpolation from lead beam data to sulfur beam data 
is desirable. 

Our study is not a substitute for a detailed dynamical
analysis. It is rather meant as a baseline calculation, designed to
point out the underlying and unifying features of the experimental
data, and to eventually extract a simpler physical message. We 
encourage dynamical microscopic prescriptions to attempt the global
study performed here. 

With respect to the recent heavy-ion experiments at CERN-SPS with
lower beam energies (40 AGeV) an interesting question is whether
the featureless $q \bar q$ spectrum is compatible with the upcoming
data.
 
\subsubsection*{Acknowledgments}
Stimulating discussions with 
W. Cassing,
L. Capelli,
O. Drapier,
A. Drees,
V. Koch,
M. Mazera,
R. Rapp,
E. Scomparin, and
G. Zinovjev are gratefully
acknowledged.
O.P.P. thanks for the warm hospitality of the nuclear theory group
in the Research Center Rossendorf.
The work is supported by 
BMBF 06DR921,
WTZ UKR-008-98 and 
STCU-015.

\begin{appendix}

\section{\label{Appendix A} Background contributions} 

Since we compose the dilepton spectra of incoherently added
thermal and background contributions one has carefully to check
the reliability of the background estimates. We employ here the
event generator PYTHIA \cite{PYTHIA} (version 6.104)
for $pp$ collisions and scale
the results to $AB$ collisions. Particular care has to be taken with
the K factors. We use the parton distribution function set
MRS D-'.

\subsection{\label{Appendix A1} Drell-Yan} 

\begin{figure}[t] 
\centering
~\\[-.1cm]
\psfig{file=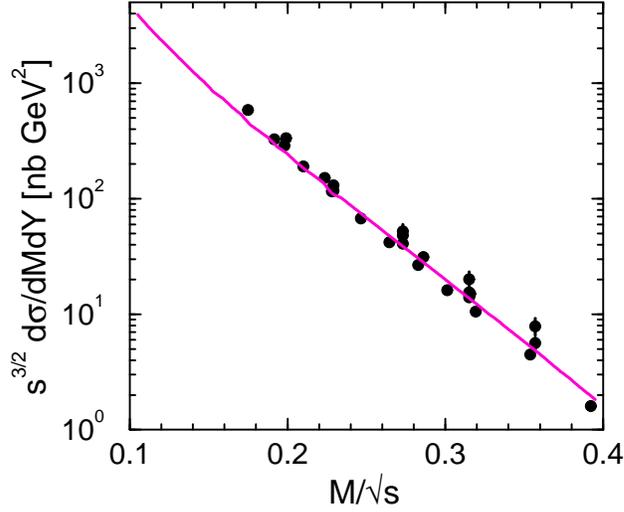,width=7cm,angle=-90}  
~\\[.5cm]
\caption{
A comparison of the Drell-Yan cross section delivered by PYTHIA
(with K factor as described in text) with the data \cite{DY_K_factor}.
}
\label{f_12}
\end{figure}

In Fig.~\ref{f_12} a data compilation of the Drell-Yan cross sections 
\cite{DY_K_factor} as
a function of the scaling variable $M/\sqrt{s}$ and our PYTHIA
simulations are displayed. The appropriate K factor
is ${\cal K}_{\rm DY} = 1.2$.
Notice, however, that these data samples are for different beam
energies and display a slight breaking of the anticipated scaling. 
Instead, independent fits of the invariant mass distributions of
the data from \cite{DY_K_factor,DY_K_factor_2} deliver an averaged
K factor of 1.5.

\begin{figure}[t] 
\centering
~\\[-.1cm]
\psfig{file=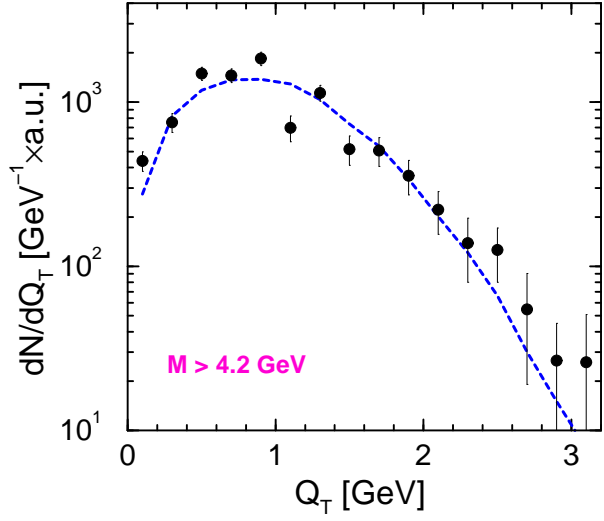,width=7cm,angle=-90}
~\\[.5cm]
\caption{
Unnormalized transverse momentum distribution of dileptons in the Drell-Yan
region, $M > 4.2$ GeV, in the NA38 experiment \protect\cite{NA38}. 
The solid curve is our
result from PYTHIA simulations with 
$\langle k_\perp^2 \rangle = (0.8 \, \mbox{GeV})^2$.
A value of $\langle k_\perp^2 \rangle = (1 \, \mbox{GeV})^2$,
as implemented in the recent PYTHIA version 6.143, results in an apparently
worse description of the data.
}
\label{f_4}
\end{figure}

The intrinsic transverse parton momentum distribution can be fixed
by the dilepton $Q_\perp$ spectrum in the Drell-Yan region
$M > 4.2$ GeV. A comparison with the NA38 data \cite{NA38}
delivers a value of $\langle k_\perp^2 \rangle = (0.8 \, \mbox{GeV})^2$
(see Fig.~\ref{f_4}).
We use in all calculations (also for charm) this value.
The K factor is less affected by variations of  
$\langle k_\perp^2 \rangle$.
 
\subsection{\label{Appendix A2} Open charm} 

\begin{figure}[t] 
\centering
~\\[-.1cm]
\psfig{file=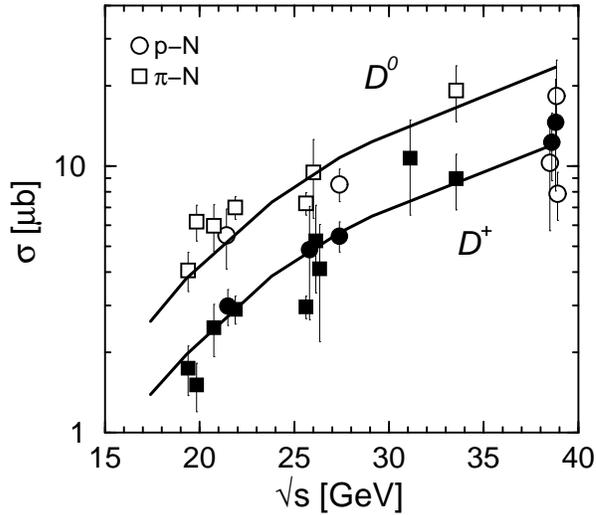,width=7cm,angle=-90}  
~\\[.5cm]
\caption{
A comparison of the open charm cross section (data compilation from
\cite{PBM}) with our PYTHIA calculations (with K factors
as described in text).
}
\label{f_13}
\end{figure}

\begin{figure}[t] 
\centering
~\\[-.1cm]
\psfig{file=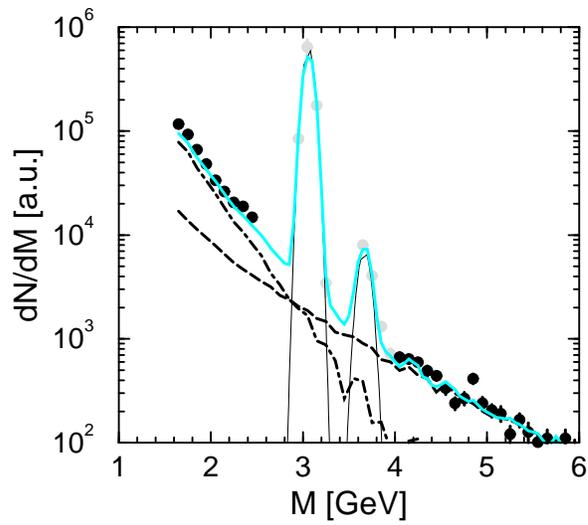,width=7cm,angle=-90} 
~\\[.5cm]
\caption{
A comparison of our PYTHIA calculations with the dilepton
spectrum in p(450 GeV) + W reactions \protect\cite{pW_Capelli}.
Dashed curve: Drell-Yan,
dot-dashed curve: open charm contribution, 
gray curve: sum of all contributions.
Included are parametrizations of the $J/\psi$ and $\psi'$
contributions according to \protect\cite{pW_Capelli}.
}
\label{f_14}
\end{figure}

We use here PYTHIA with charm mass parameter $m_c = 1.5$ GeV
and default fragmentation ''hybrid''.
There are two different sources of a determination of the open charm
K factor. Either one uses the hadronic channels, i.e.\ cross sections
of identified $D^0$ and $D^+$ mesons from Fermi lab experiments,
which are compiled in \cite{PBM}, or the NA50  dilepton
data \cite{pW_Capelli} for the reaction p(450 GeV) + W.
Fig.~\ref{f_13} shows the comparison with hadron channels,
which deliver K factors of ${\cal K}_{D^0} = 5.4$ and 
${\cal K}_{D^+} = 6.8$ resulting in an averaged open charm K factor
of ${\cal K}_c = 5.7$. Obviously these data do not constrain the
K factor very reliably. A sharper constraint is given by the
dilepton channel which confirms this value
as seen in Fig.~\ref{f_14}. 
The uncomfortably large K factor points to higher order processes
which, however, could change the spectral shapes.  
In this respect the envisaged experiments by the NA60 collaboration 
\cite{NA6i} to
identify explicitly open charm are very important for this 
dilepton source.  
\end{appendix}

\newpage

\end{document}